\documentclass[fleqn,usenatbib]{mnras}

\usepackage[T1]{fontenc}
\usepackage[british]{babel}
\usepackage{amsmath,amssymb}
\usepackage{newtxtext,newtxmath}

\usepackage{graphicx,pgf}
\usepackage[labelformat=simple]{subcaption}

\usepackage{adjustbox}
\usepackage{import}

\usepackage{ulem}
\usepackage{siunitx}
\usepackage{hyperref}
\hypersetup{colorlinks = true}
\usepackage{orcidlink}

\usepackage{xspace}
\usepackage{flushend}

\usepackage{xpatch}
\makeatletter
\xpatchcmd\NAT@citex
 {%
  \@citea\NAT@hyper@{%
    \NAT@nmfmt{\NAT@nm}%
    \hyper@natlinkbreak{\NAT@aysep\NAT@spacechar}{\@citeb\@extra@b@citeb}%
    \NAT@date
  }%
 }
 {%
  \@citea
  \NAT@nmfmt{\NAT@nm}%
  \NAT@aysep\NAT@spacechar
  \NAT@hyper@{\NAT@date}%
 }
 {}{}
\xpatchcmd\NAT@citex
 {%
  \@citea\NAT@hyper@{%
    \NAT@nmfmt{\NAT@nm}%
    \hyper@natlinkbreak{\NAT@spacechar\NAT@@open\if*#1*\else#1\NAT@spacechar\fi}%
    {\@citeb\@extra@b@citeb}%
    \NAT@date
  }%
 }
 {
  \@citea
    \NAT@nmfmt{\NAT@nm}%
    \NAT@spacechar\NAT@@open\if*#1*\else#1\NAT@spacechar\fi
    \NAT@hyper@{\NAT@date}%
 }
 {}{}
\makeatother

\title[Modelling lensing in the eclipse of \psr]{Discovery and modelling of broad-scale plasma lensing in black-widow pulsar \mbox{J2051\,$-$\,0827}} 

\author[F. X. Lin et al.]{F. X. Lin$^{\orcidlink{0000-0002-6820-4275}}$\!,$^{1,2}$\thanks{E-mail:flin@cita.utoronto.ca} R. A. Main$^{\orcidlink{0000-0002-7164-9507}}$\!,$^{3}$  J. P. W.~Verbiest$^{\orcidlink{0000-0002-4088-896X}}$\!,$^{4,3}$, M. Kramer$^{\orcidlink{0000-0002-4175-2271}\,3,5}$ and G. Shaifullah$^{\orcidlink{0000-0002-8452-4834}\,6,7}$\\
$^{1}$Canadian Institute for Theoretical Astrophysics, University of Toronto, 60 St George St., Toronto, ON M5S 3H8 , Canada \\
$^{2}$Department of Physics, University of Toronto, 60 St George St., Toronto, ON M5S 1A7, Canada\\
$^{3}$Max-Planck-Institut f{\"u}r Radioastronomie, Auf dem H{\"u}gel 69, D-53121 Bonn, Germany \\
$^{4}$Fakult{\"a}t f{\"u}r Physik, Universit{\"a}t Bielefeld, Postfach 100131, D-33501 Bielefeld, Germany \\
$^{5}$Jodrell Bank Centre for Astrophysics, School of Physics and Astronomy, The University of Manchester, Manchester M13 9PL, UK \\
$^{6}$ ASTRON, the Netherlands Institute for Radio Astronomy, Postbus 2, NL-7900 AA, Dwingeloo, The Netherlands\\
$^{7}$ Dipartimento di Fisica `G. Occhialini', Universit\'{a} degli Studi di Milano-Bicocca, Piazza della Scienza 3, I-20126 Milano, Italy\\
}

\begin{document}

\newcommand{\msun}{\mbox{M$_{\odot}$}}
\newcommand{\rsun}{\mbox{R$_{\odot}$}}
\newcommand{\aopx}{\mbox{$\Delta_{\pi \rm M}$}}
\newcommand{\shap}{\mbox{$\Delta_{\rm S}$}}
\newcommand{\psr}{\mbox{PSR~J2051\,$-$\,0827}\xspace}

\definecolor{midori}{rgb}{0.165,0.376,0.231}
\newcommand{\RM}[1]{\textcolor{purple}{RM: #1}}
\newcommand{\FX}[1]{\textcolor{blue}{FX: #1}}
\newcommand{\FXdone}[1]{\textcolor{blue}{\sout{FX: #1}}}
\newcommand{\GS}[1]{\textcolor{midori}{GS: #1}}
\newcommand{\changed}[1]{\textcolor{orange}{#1}}

\newcommand{\dphi}{\delta\!\varphi}
\newcommand{\phig}{\phi_\text{geo}}
\newcommand{\phid}{\phi_\text{disp}}
\newcommand{\phit}{\phi_\text{tot}}
\newcommand{\veff}{v_\text{eff}}
\newcommand{\vrel}{v_\text{rel}}
\newcommand{\dpsr}{d_\text{psr}}
\newcommand{\dlens}{d_\text{lens}}
\newcommand{\xpsr}{x_\text{psr}}
\newcommand{\xobs}{x_\text{obs}}
\newcommand{\ppsr}{P_\text{psr}}
\newcommand{\DM}{D\!M}
\renewcommand{\us}{\si{\us}\xspace}

\newcommand{\pb}{\mbox{$P_{\rm b}$}}
\newcommand{\pbdot}{\mbox{$\dot{\pb}$}}

\newcommand{\wvect[1]}{{\bsf #1}}
\newcommand{\bvect[1]}{{\bsf #1}$_{\bm 0}$}

\newcommand{\msout}[1]{\text{\sout{\ensuremath{#1}}}}

\let\oldhat\hat
\renewcommand{\vec}[1]{\mathbf{#1}}
\renewcommand{\hat}[1]{\oldhat{\mathbf{#1}}}

\date{Accepted 2021 June 22. Received 2021 June 22; in original form 2020 October 2}

\pagerange{\pageref{firstpage}--\pageref{lastpage}} 
\pubyear{2021}

\maketitle

\label{firstpage}
  
\begin{abstract}
We report on an unusually bright observation of \psr recorded during a regular monitoring campaign of black-widow pulsar systems with the Effelsberg 100-m telescope. Through fortunate coincidence, a particularly bright scintillation maximum is simultaneous with the eclipse by the companion, enabling precise measurements of variations in the flux density, dispersion measure (DM), and scattering strength throughout the eclipse.  The flux density is highly variable throughout the eclipse, with a peak 1.7 times the average away from the eclipse, and yet does not significantly decrease on average. 
We recover the flux density variations from the measured DM variations using geometric optics, with a relative velocity as the only free parameter. 
We measure an effective velocity of \SI[separate-uncertainty=true]{470 \pm 10}{km.s^{-1}}, consistent with the relative orbital motion of the companion, suggesting that the outflow velocity of the lensing material is low, or is directly along the line of sight. 
The two percent uncertainty on the effective velocity is a formal error; systematics related to our current model are likely to dominate, and we detail several extensions to the model to be considered in a full treatment of lensing. 
This is a demonstration of the causal link between DM and lensing; the flux density variations can be predicted directly through the derivatives of DM. 
Going forward, this approach can be applied to investigate the dynamics of other eclipsing systems, and to investigate the physical nature of scintillation and lensing in the ionized interstellar medium. 

\end{abstract}

\begin{keywords}
stars: atmospheres -- binaries: eclipsing -- pulsars: general -- pulsars: individual: PSR~J2051~--~0827.
\end{keywords}

\section{Introduction}
\label{sec:intro}
Despite decades of studies, neutron stars are still providing researchers with unique opportunities to study a broad spectrum of physics. This is partially due to a plethora of ways in which neutron stars manifest themselves observationally and the variety of environments in which we find them \citep[see][for an overview]{kaspi10}. With about 2800 neutron stars currently detectable in radio \citep[][]{manchester+05}\footnote{\href{http://www.atnf.csiro.au/research/pulsar/psrcat}{http://www.atnf.csiro.au/research/pulsar/psrcat V1.63}}, we can probe populations of rarely occurring neutron stars in exotic stellar systems. An example of such a class are the so-called ``spiders'', pulsars in tight, often eclipsing, binaries in which the low-mass companion star is ablated by the emission from the neutron star \citep[first discovered by][]{fruchter+88}. In this work we focus on \psr (discovered by \citealt{stappers+96}), a member of the so-called `black widows'. 

Spiders are an important laboratory for interesting physics; although many eclipse mechanisms have been proposed (see \citealt{thompson+94} for a detailed review), there is no seeming consensus on the correct eclipse mechanism, and some systems needing to invoke multiple mechanisms across frequency to agree with observations \citep{fruchter+92, polzin+20}.  
Importantly, black widows may explain the origin of isolated millisecond pulsars (MSPs, first discovered by \citealt{backer+82}). 
Such pulsars are expected to be formed via accretion of matter from a stellar companion, and the ablation of a companion after accretion is possibly a viable way to create such a pulsar in isolation \citep[e.g.][]{ruderman+89}. 
Long term observations of \psr reveal significant orbital variability of the system \citep{shaifullah+16}, and inferred mass-loss rates suggest it may evaporate within a Hubble time \citep{stappers+96,stappersOrbitalEvolutionProper1998,polzin+19}.

Recently, extreme plasma lensing has been seen surrounding eclipses in two eclipsing pulsars, the original Black Widow pulsar B1957\,$+$\,20 \citep{main+18, li+19}, and Redback system B1744\,$-$\,24A \citep{bilous+19}.  The effects of lensing were seen through highly magnified pulses, amplified by factors of $\gtrsim 10$ over tens of ms.  The lensing appears to resolve the pulse emission, constraining emission sizes, and separations, and can be used to infer the velocity of the eclipsing outflow; however, it is difficult to measure these physical properties quantitatively, as it is very difficult to measure the dispersion measure (DM) precisely on such small time-scales, depending on the pulse structure and its single pulse S/N. As such, constraints were obtained with a simplistic approach, using the measured magnifications, time, and frequency scales of lensing to infer lens sizes and velocities.  In a region of higher column density, \citet{main+18} presented evidence of less extreme lensing, with $\gtrsim \SI{100}{ms}$ time-scales and magnifications of $\sim2$, but could not draw any quantitative constraints from this region.

In this publication, we present an analysis of a single epoch observation of \psr that covers two eclipses and shows the first clear evidence of plasma lensing in this system.  During the first of these eclipses a scintillation maximum occurred partly overlapping with the eclipse, allowing us to study the eclipse in unprecedented detail. The paper is organized as follows:  In Section~\ref{sec:data}, we report on the observations and data processing. In Section~\ref{sec:measurements}, we present our methods as well as the derived quantities of flux density, DM, scattering, and polarization throughout the eclipse. In Section~\ref{sec:lensing}, we present our geometric optics modelling of the flux density throughout the eclipse and argue that the flux density variations are caused by plasma lensing. In Section~\ref{sec:physics}, we discuss the physical constraints and properties of the eclipsing material. In Section~\ref{sec:limitations}, we discuss some simplifying assumptions and limitations of our model, and we conclude and summarize our findings in Section~\ref{sec:summary}.

\section{Observations}
\label{sec:data}

Observations of \psr\ analysed in this work were recorded on 2012 September 16 using the 100-m Effelsberg radio telescope and the central beam of a seven-beam receiver `P217mm'\footnote{\href{https://eff100mwiki.mpifr-bonn.mpg.de/doku.php?id=information_for_astronomers:rx:p217mm}{https://eff100mwiki.mpifr-bonn.mpg.de/}}. The data from two circularly polarized dipoles were digitized to 8-bit values using the Field-programmable gate array (FPGA)- and off-the-shelf server-based coherent de-dispersion frequency-multiplexing software backend \textsc{psrix} \citep{lazarus+16}. Full polarization detection was followed by folding of the observations into 1024 phase bins spanning the topocentric rotational period of the pulsar. We performed coherent de-dispersion using a convolving polyphase filter bank implemented in \textsc{dspsr} \citep{vanStraten+11} with $\DM=\SI{20.7449}{pc.cm^{-3}}$ and after full-Stokes \citep[as defined in][]{vanStraten+10} detection, the mean phase-resolved light curve of the pulsar, referred to as the pulse profile, was output every 10 seconds throughout the \SI{3.34}{h} of observation. The duration of the observation was chosen to cover more than one orbit of the pulsar. 

\begin{figure*}
    \begin{center}
        \begin{adjustbox}{clip,trim=0.3cm 1.8cm 1.2cm 1.8cm}
        \input{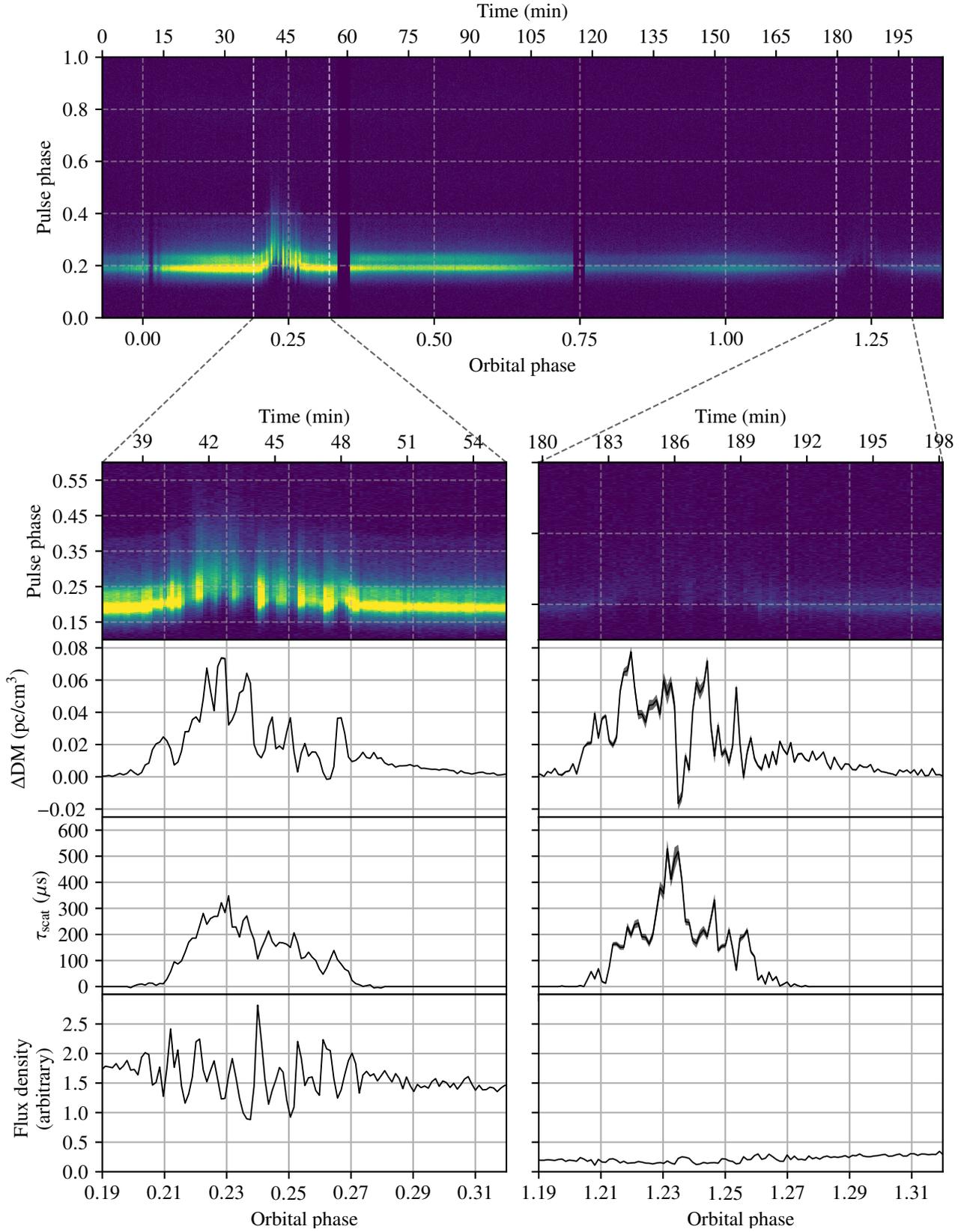}
        \end{adjustbox}
    \end{center}
    \caption{Measurements of variable DM, scattering, and flux density in two subsequent eclipses.
    Top panel: Pulse profile across \SI{3.34}{h} of observations. The dropout near \SI{10}{min} from the start of the observations is due to instrumental effects, and the subsequent drops at $\sim 60$ and $\sim \SI{115}{min}$ are from calibrator scans.
    Bottom panels: Zoomed-in profiles near the two eclipses. First row: zoom of pulse profile. Second row: DM difference to outside the eclipse. Third row: scattering time-scale $\tau_\mathrm{scat}$ at central frequency 1345.5 MHz. Bottom row: pulse-averaged flux density in arbitrary units.
    The error bars are present in the first eclipse, but are too small to see on the figure. They can be seen in the $\Delta \DM$ and $\tau_\mathrm{scat}$ of the second eclipse.
    }
    \label{fig:measurements}
\end{figure*}

We sum both orthogonal polarizations to form total intensity $I$, and we determine an off-gate as bins 500--600 corresponding to roughly pulse phase of 0.99--1.09 in Fig.~\ref{fig:measurements}, away from the visible emission and scattering tail.  We divide $I$ by the time average of the off-pulse region across the full observation, and subtract the mean of the off-pulse region in each sub-integration to roughly correct for gain variations. 
Sub-integrations with an off-pulse standard deviation $>5\times$ the mean were masked, while sub-band edges and corrupted time bins were additionally masked by hand.

\begin{figure*}
\label{fig:system_geometry}
\begin{subfigure}[t]{.49\textwidth}
  \centering
  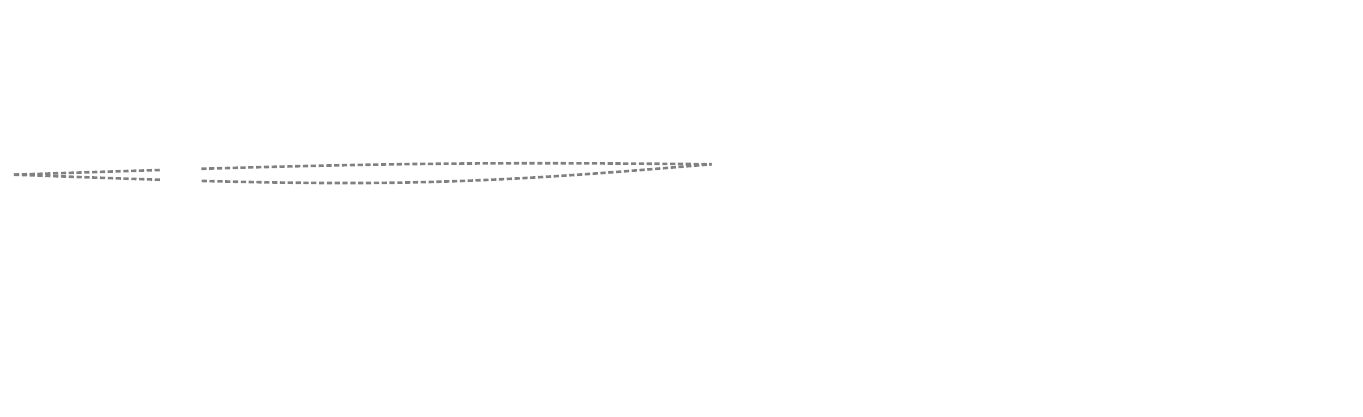
  \caption{}
  \label{fig:sub-first}
\end{subfigure}
\hfill
\begin{subfigure}[t]{.49\textwidth}
  \centering
  %% Creator: Inkscape 1.0 (4035a4fb49, 2020-05-01), www.inkscape.org
%% Accompanies image file 'system_diagram_faceon.pdf' (pdf, eps, ps)
\begingroup%
  \makeatletter%
  \providecommand\color[2][]{%
    \errmessage{(Inkscape) Color is used for the text in Inkscape, but the package 'color.sty' is not loaded}%
    \renewcommand\color[2][]{}%
  }%
  \providecommand\transparent[1]{%
    \errmessage{(Inkscape) Transparency is used (non-zero) for the text in Inkscape, but the package 'transparent.sty' is not loaded}%
    \renewcommand\transparent[1]{}%
  }%
  \providecommand\rotatebox[2]{#2}%
  \newcommand*\fsize{\dimexpr\f@size pt\relax}%
  \newcommand*\lineheight[1]{\fontsize{\fsize}{#1\fsize}\selectfont}%
  \ifx\svgwidth\undefined%
    \setlength{\unitlength}{419.98178672bp}%
    \ifx\svgscale\undefined%
      \relax%
    \else%
      \setlength{\unitlength}{\unitlength * \real{\svgscale}}%
    \fi%
  \else%
    \setlength{\unitlength}{\svgwidth}%
  \fi%
  \global\let\svgwidth\undefined%
  \global\let\svgscale\undefined%
  \makeatother%
  \begin{picture}(1,0.20468224)%
    \lineheight{1}%
    \setlength\tabcolsep{0pt}%
    \put(0,0){\includegraphics[width=\unitlength,page=1]{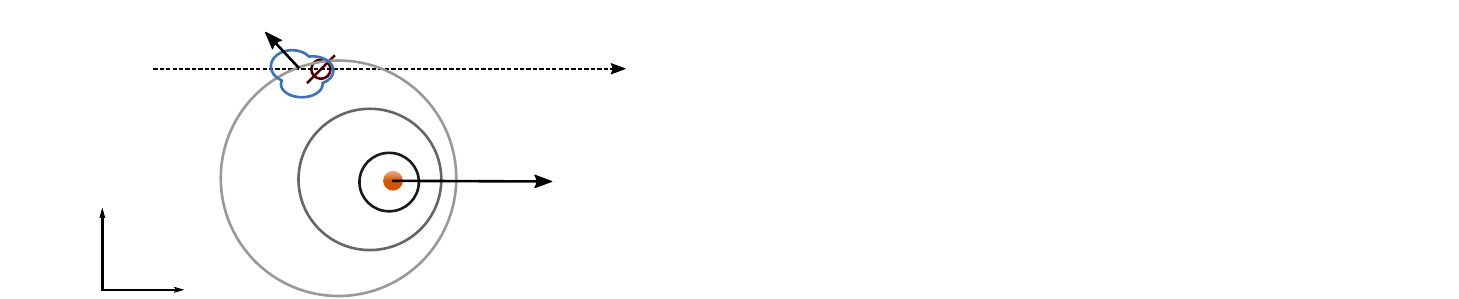}}%
    \put(-0.00197646,0.13721234){\color[rgb]{0,0,0}\makebox(0,0)[lt]{\lineheight{0.62499928}\smash{\begin{tabular}[t]{l}$DM\left(\frac{v_\text{eff}\,t}{R_F}\right) = DM(x)$\end{tabular}}}}%
    \put(0.14490163,0.18659184){\color[rgb]{0,0,0}\makebox(0,0)[lt]{\lineheight{0.62500107}\smash{\begin{tabular}[t]{l}$v_\text{flow}$\end{tabular}}}}%
    \put(0.30904242,0.05051536){\color[rgb]{0,0,0}\makebox(0,0)[lt]{\lineheight{0.62500107}\smash{\begin{tabular}[t]{l}$v_\text{orb}=\frac{K_p}{\sin i}\frac{m_p+m_c}{m_c}$\end{tabular}}}}%
    \put(0.12871249,0.00209949){\color[rgb]{0,0,0}\makebox(0,0)[lt]{\lineheight{0.62500107}\smash{\begin{tabular}[t]{l}$x$\end{tabular}}}}%
    \put(0.06495428,0.06787462){\color[rgb]{0,0,0}\makebox(0,0)[lt]{\lineheight{0.62500107}\smash{\begin{tabular}[t]{l}$y$\end{tabular}}}}%
  \end{picture}%
\endgroup%
 
  \caption{}
  \label{fig:sub-second}
\end{subfigure}
\caption{Schematics of the \psr binary system and the lensing set-up, drawn at the orbital phase of $\sim$ 0.28. The companion (the orange sphere), the binary separation $a_\text{orb}\approx 0.72 R_{\odot}$ and inclination $\sim 40^\circ$, and the eclipse size $\sim 10$ per cent of the orbit are drawn to scale using sizes from \citet{stappersLightCurveCompanion1998} and \citet{polzin+19}. The blue `cloud' shows a lensing blob of underdense plasma. See Section~\ref{sec:physics} for a detailed discussion of pictured velocities. 
(a)~Edge-on view. The ISM lens is not shown. The asymmetry of the outflowing material can be seen in the DM time-series in Fig.~\ref{fig:measurements}, and is similar to that seen in other black widows. The orbital velocity of the companion is roughly in the $x$-axis, out of the paper. 
(b)~Face-on view. In this frame, the pulsar is thought of as being stationary, and the companion and its outflow moving across our line of sight. The dashed line indicates the axis along which DM is sampled in the eclipse. The concentric circles roughly indicate contours of electron density. The largest circle is to scale with the size of the eclipse.
}
\end{figure*}

\section{Measurements}
\label{sec:measurements}

In this section, we discuss the 2D template matching method used to measure the DM, scattering time-scales, and flux density. 
Note that the DM we refer to throughout the rest of this paper is relative to the nominal $\DM=\SI{20.7449}{pc.cm^{-3}}$ that the pulse profiles are de-dispersed to. 
We further discuss disentangling the interstellar scintillation (ISS) from the flux density variations due to the eclipsing material, measuring the magnification curve and spectrum throughout the eclipse. 

\subsection{2D template matching -- variable dispersion measure and scattering}
In addition to the flux density varying significantly over the eclipse, many \SI{10}{s} subintegrations also appear to be dispersed and scattered. 
We employ a phase-frequency domain template matching method to simultaneously measure dispersion, scattering, and flux density. 
Our implementation\footnote{\href{https://github.com/quantumfx/dmfitter}{https://github.com/quantumfx/dmfitter}} is independent, but similar, to that introduced by \citet{pennucciElementaryWidebandTiming2014}, which is an extension of the classical Fourier domain template matching method from \citet{taylorPulsarTimingRelativistic1992} to allow for a frequency-resolved template. 

As noted by \citet{pennucciElementaryWidebandTiming2014}, there is a large amount of freedom in choosing the time--frequency template profile $p(\varphi,\nu)$. 
Since there appears to be little profile evolution across our band outside the eclipse, we take the time- and frequency-averaged profile in the time window 15--35 min, where the signal is brightest (see Fig.~\ref{fig:measurements}) as our template, so that $p(\varphi,\nu)$ is frequency independent. 
In this section, we briefly describe the phase-frequency template matching algorithm, and defer the extra details of the implementation to \citet{pennucciElementaryWidebandTiming2014} or our publicly available code.

Suppose the pulsar has some intrinsic phase-frequency profile $p(\varphi,\nu)$. The observed data $d(\varphi,\nu)$ can be described as
\begin{align}
    d(\varphi,\nu) = I(\nu)\,\big(p\left(\varphi-\dphi(\nu),\nu\right)*g(\varphi,\nu,\tau)\big)+b(\nu)+n(\varphi,\nu)
\end{align}
where $I(\nu)$ describes the amplitude of the frequency channels, describing e.g. the scintillation; $*$ denotes a convolution in $\varphi$; $g(\varphi,\nu,\tau)$ a normalized exponential scattering tail with scattering time-scale $\tau$; $b(\nu)$ a frequency-dependent offset term, essentially describing the shape of the bandpass; $n(\varphi,\nu)$ a additive noise term, typically assumed to be Gaussian with zero mean; and $\dphi(\nu)$ a frequency-dependent shift term. 
We assume the extra scattering to scales in frequency as $\nu^{-4}$, but note that the exponent can be an additional parameter of the fit. 
The frequency-dependent shift $\dphi_k=\dphi(\nu_k)$ will be due to dispersion, $k$ indexing the frequency channels. Thus, following \citet{pennucciElementaryWidebandTiming2014}, we can write 
\begin{align}
    \label{eq:dphik}
    \dphi_k = \frac{k_\text{DM}\,\DM}{\ppsr}\nu_k^{-2}
\end{align}  
where we set the reference phase to $0$, so that the DM is measured relative to the out-of-eclipse region, and take 
\begin{align}k_\text{DM} \equiv c r_\text{e}/2\pi = \SI{4.148\,806\,4239\,(11)}{GHz^2.ms.cm^3.pc^{-1}}
\end{align} with $c$ the speed of light and $r_\text{e}$ the classical electron radius \citep[see e.g.][]{kulkarniDispersionMeasureConfusion2020}.
Parameters $I(\nu)$, DM, and $\tau$ can then be estimated by standard $\chi^2$ minimization, and their associated errors from inverting the Hessian of $\chi^2/2$ at the minima. 

The $\DM$ and $\tau$ throughout our observations give the DMs $\DM(t)$, the scattering time-scales $\tau(t)$, and the amplitudes $I(t,\nu)$ as functions of time.  
We use $I(t,\nu)$ as the dynamic spectrum throughout; it represents the measured intensity, accounting for the scattered flux density.
When both $\DM$ and $\tau$ are close to 0 (i.e. outside the eclipse), they are highly contravariant since increasing one and decreasing the other nearly cancels each other out in the resulting frequency-dependent time delay. 
The minimizer attempts to correct for the minor shape differences between the data and the template using the two parameters, leading to incorrect results where $\DM\propto -\tau$. 
To account for this, we note that the rise in $\tau$ is relatively sharp at the eclipse (see Fig.~\ref{fig:measurements}), thus we fit for both $\DM$ and $\tau$ where scattering is significant, in the orbital phase range 0.20--0.28, and set $\tau$ to \SI{0}{\us} elsewhere. 

\subsection{Determining magnifications -- removing the effects of ISM scintillation}
\label{sec:measure_magnification}
A complicating factor to modelling lensing from the eclipsing wind is the effect of ISS.  While we can measure the relative flux density very precisely,  what we wish to observe is rather the magnification $\mu = I / \langle I \rangle$, the ratio of the intensity to the average.  To measure magnification effects solely from the eclipsing material, we must separate it from the effects of ISS.  Under the assumption that the pulsar appears suitably point-like after being scattered in the eclipse (see Section~\ref{sec:eclipsescreen}), the lensing contributions from the eclipse and ISS are multiplied, 
\begin{equation}
    I_{\rm obs}(t, \nu) = I_{\rm int}(t) \mu_{\rm ISS}(t, \nu) \mu_{\rm ecl}(t, \nu).
\end{equation}
We assume that $\langle I_{\rm int}(t)\rangle$ is stable over time, so we need only to measure $\mu_{\rm ISS}(t, \nu)$, and account the remaining pulse-to-pulse variations as an additional error in the magnification measurements. 
Moreover, any intrinsic spectral index and frequency variation in $\langle I_\text{int}(t)\rangle$ are absorbed into the magnification terms.
Outside the eclipse, the intrinsic variations, even averaged over 10\,s ($=$ 2218 pulse rotations), result in 2 per cent RMS variations, much larger than the measurement errors of the frequency-averaged intensity $I(t)$.  We add this in quadrature to the error on $\mu(t)$.
 
The time-scale for ISS is much larger than the flux density variations during eclipse, making them separable (which can be seen in the top panel of Fig.~\ref{fig:measurements}). 
We Wiener filter and inpaint $I_\text{obs}(t,\nu)$ with the eclipse region masked to get an estimate for $\mu_\text{ISS}$, as described in the appendix. We then divide the Wiener filtered $I_\text{obs}(t,\nu)$ by $\mu_\text{ISS}$ without the eclipse region masked to get the full frequency-resolved magnification $\mu_{\rm ecl}(t, \nu)$. 
Short time-scale RMS variations of order 2 per cent still exists in the Wiener-inpainted spectrum, as seen in the top right-hand panel of Fig.~\ref{fig:wiener_dynspec}, which may bias the estimation of $\mu_{\rm ecl}(t, \nu)$. 
While it is beyond the scope of this paper, future studies of this type can generate ensembles of constrained realizations of $\mu_\text{ISS}$ to study the effect of such bias.

\begin{figure*}
    \begin{center}
        \begin{adjustbox}{clip,trim=0.3cm 0cm 1.3cm 0cm}
        \input{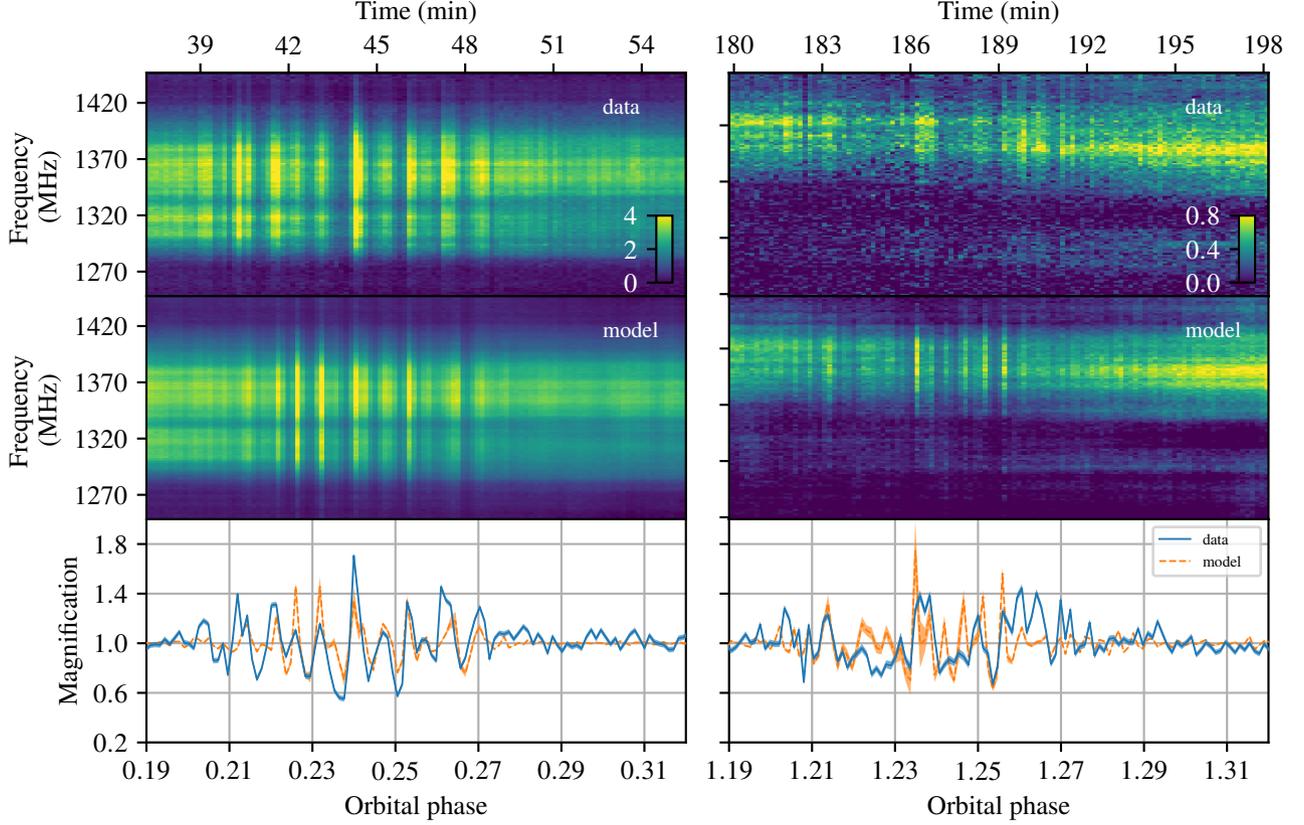}
        \end{adjustbox}
    \end{center}
    \vspace{-5mm}
    \caption{
    Fits of our one-parameter model to the dynamic spectra. The left-hand and right-hand panels show the first and second eclipses, respectively. Top row: Measured Wiener-filtered dynamic spectra. Middle row: Product of the interstellar scintillation spectra and the computed magnification spectra. Bottom row: Measured (the blue, solid curves) and model magnifications ( theorange, dashed curves). The effective velocity $v_{\rm eff} = \SI{470}{km.s^{-1}}$ is fit for the first eclipse, then applied for the second eclipse in our model. 
    }
    \label{fig:fit}
\end{figure*}

\begin{figure}
    \begin{center}
        \begin{adjustbox}{clip,trim=0.8cm 0.0cm 0.7cm 0.45cm}
        \input{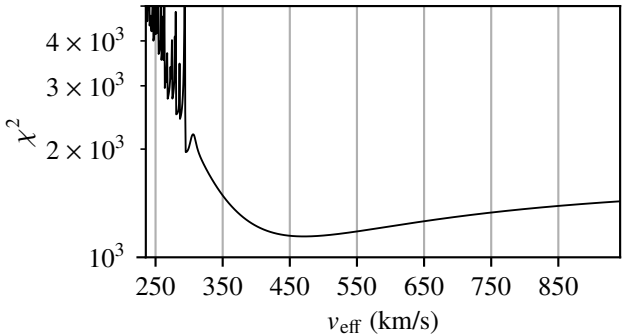}
        \end{adjustbox}
    \end{center}
    \vspace{-5mm}
    \caption{
    $\chi^2$ versus effective velocity for the first eclipse using 90 data points from the frequency-averaged magnification between orbital phase 0.187 and 0.293. The high variability below $v_{\rm eff} \lesssim \SI{300}{km.s^{-1}}$ is due to the appearance of caustics in our band in the model (see Fig.~\ref{fig:2Dmodel}).
    }
    \label{fig:veff}
\end{figure}

\section{Modelling lensing from dispersion measure}
\label{sec:lensing}
\subsection{Geometrical optics of plasma lenses}
As a pulsar moves behind a variable electron column, we measure the DM variations along the axis of motion (See Fig.~\ref{fig:sub-first} for the lensing system). 
Since we have little to no information about the direction transverse to the source motion, we proceed by assuming the DM is a 1D, thin lens \citep[see e.g.][]{cleggGaussianPlasmaLens1998,cordesLensingFastRadio2017}, and discuss the effects of this assumption in Section~\ref{sec:limitations_screendim}. 
A ray passing through a region of excess DM will have its phase changed by an amount
\begin{equation}
    \label{eq:phidisp}
    \phid(x,\lambda) = - \lambda r_\text{e} \DM(x)
\end{equation}
where $\lambda$ is the wavelength of the ray, $r_\text{e}$ the classical electron radius, and $x$ the position on the image plane. Since additional DM advances phase, the local maxima of DM are divergent, and the local minima convergent. 
This can qualitatively be seen in Fig.~\ref{fig:measurements}; the brightest times in the eclipse correspond to minima in the DM curve, and vice-versa.
In this section, we formalize this relationship. 

Rays of light will travel by paths of extremal phase, which in the absence of a lens is the direct line of sight. 
Off the line of sight, the geometric contribution to the phase for a ray is 
\begin{align}
    \phig(x,&\xpsr,\xobs,\lambda) = \frac{\pi}{ R_\text{F}^2}\bigg(\frac{\dlens}{\dpsr}(x-\xpsr)^2 \nonumber+\\& \frac{\dpsr-\dlens}{\dpsr}(x-\xobs)^2 - \frac{(\dpsr-\dlens)\,\dlens}{\dpsr^2}(\xobs-\xpsr)^2\bigg )
\end{align}
where $R_\text{F}\equiv\sqrt{\lambda (\dpsr-\dlens)\,\dlens/\dpsr}$ is the Fresnel scale. Since we are in the limit $d_\text{pl} \equiv \dpsr-\dlens \ll \dpsr$, the Fresnel scale reduces to $R_\text{F}\approx\sqrt{\lambda d_\text{pl}}$, and the geometric phase reduces to
\begin{align}
    \phig(x,\xpsr,\lambda) \approx \frac{\pi}{R_\text{F}^2}(x-\xpsr)^2
\end{align}
independent of the observer position. The total phase of the ray is then the total of the geometric and dispersive contributions
\begin{align}
    \phi_\text{tot}(x,\xpsr,\lambda) \approx \frac{\pi}{R_\text{F}^2}(x-\xpsr)^2 - \lambda r_\text{e} \DM(x)
\end{align}
where the negative sign comes from the phase velocity being faster than $c$.
Images occur at extrema of phase $\partial_x\phit(x,\xpsr)=0$, which is the lens equation. 
A more conventional way of writing it, relating the source and image positions, is
\begin{align}
    \label{eq:lenseq}
    \xpsr = x - \frac{\lambda r_\text{e} R_\text{F}^2}{2\pi}\partial_x \DM(x).
\end{align}
In the absence of $\phid$, $\phit$ has one minimum at $x=\xpsr$ from $\phig$. Adding a smooth $\phid$ can create extrema in pairs of maxima and minima $\phit$. This is an example of the \textit{odd-image theorem}. 
For any given source position $\xpsr$, there are thus an odd number of real solutions $x_i, i=1,2...,n$ to the lens equation~(\ref{eq:lenseq}), which are the locations of the images on the lens plane. 

\begin{figure}
    \begin{center}
        \begin{adjustbox}{clip,trim=0.65cm 1.5cm 0.7cm 2.8cm}
        \input{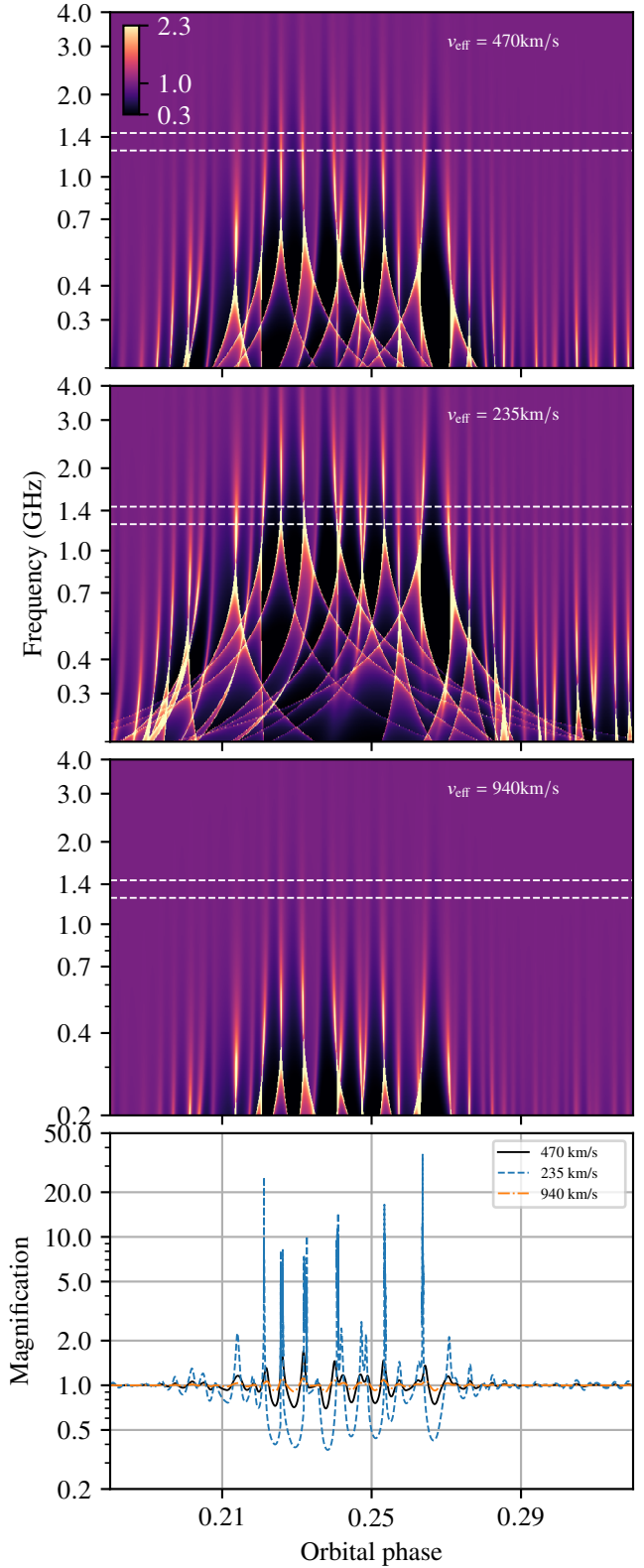}
        \end{adjustbox}
    \end{center}
    \vspace{-5mm}
    \caption{
    Images show predicted magnifications for range of frequencies from 200 to \SI{4000}{MHz} for three different effective velocities, our best-fitting velocity of \SI{470}{km.s^{-1}} (\textit{top}), and with \SI{235}{km.s^{-1}} (\textit{middle}), and \SI{940}{km.s^{-1}} (\textit{bottom}) for reference. The white-dashed lines show the extent of our band.  Bottom panel shows the average magnification for each of the three velocities in our band. At lower velocities, significantly more caustics are formed, and there is predicted power below the \SI{10}{s} integrations of our data.
    }
    \label{fig:2Dmodel}
\end{figure}

Since the images $x_i$ are generally off the line of sight, they each have an associated group delay from the combined contribution of the geometric and dispersive delays
\begin{align}
\label{eq:groupdelay}
    \tau_{\text{group},\,i} = \frac{\lambda}{2\pi c}\left[\frac{\pi}{R_\text{F}^2}(x-\xpsr)^2 + \lambda r_\text{e} \DM(x)\right]_{x=x_i}.
\end{align}
The variations in DM can focus and defocus rays, leading to different images $x_i$ having different magnifications $\mu_i$. 
Since lensing is surface brightness conserving, 
magnification is given by the (differential) ratio of the lensed image size to the unlensed image size 
\begin{align}
    \label{eq:magnification}
    \mu_i &= \frac{\partial x}{\partial \xpsr}\bigg |_{x=x_i}\nonumber\\
    &= \left(1-\frac{\lambda r_\text{e} R_\text{F}^2}{2\pi}\partial^2_x \DM(x)\right)^{-1}\bigg |_{x=x_i}.
\end{align}
Thus, the magnification $\mu_i$ of an image $x_i$ at a particular source position $\xpsr$ is related to the second derivative of the DM, evaluated at that image. 
When pairs of images are created, $\mu_i$ diverge at the image locations for a point source. 
These diverging features are known as \textit{caustics}. Fig. \ref{fig:2Dmodel} shows where caustics occur in our model of the first eclipse in time and frequency.
The divergences of magnification at caustics are usually tempered by finite source-size effects, or in our case, more likely by the scattered images. We discuss caustics further in Sections~\ref{sec:limitations_geomoptics} and \ref{sec:eclipsescreen}.

In our model, we explicitly use classical geometrical optics, so the total magnification is an \textit{incoherent sum} of all the images
\begin{align}
    \label{eq:mugeom}
    \mu_\text{tot}=\sum_i |\mu_i|,
\end{align}
At lower frequencies, away from the caustics, multiple imaging can lead to fine scale interference in time and frequency between images. 
On average, the interference patterns are well approximated by the magnification predicted by geometric optics.
We do not observe such patterns in the data, and if they did exist in our observations, they are likely to have been averaged out by the coarse data resolution. 
As such, geometric optics is good enough for the purposes of this paper, which is to show a correlation between DM and light curve of the pulsar across the eclipse. 

\subsection{Magnifications from DM measurements}
\label{sec:magfromdm}
To compute magnification from the measured DM, a spatial model of the lens is required. 
This would be the physical column density of electrons. 
However, we can only measure DM, which is a priori affected by lensing effects such as multipath propagations. 
In other words, we require DM as a function of lens coordinate $x$, but we only measure DM as a function of the source coordinate $x_\text{psr}$, related to each other by the lens equation~(\ref{eq:lenseq}).
Undoing propagation in the measured DM is a difficult problem beyond the scope of this paper. 
In our modelling, we take measured DM to be the physical column, i.e. approximate $x_\text{psr}(t)\simeq x$ for the computation of the derivatives of the DM. 
This introduces a bias in regions with DM gradients due to the line of sight being shifted. 
To account for this, consider
\begin{align}
\label{eq:dmerr}
    \DM(\xpsr) &= \DM\left(x-\frac{\lambda r_\text{e} R_\text{F}^2}{2\pi}\partial_x \DM(x)\right)\nonumber\\
            &\simeq \DM(x) - \frac{\lambda r_\text{e} R_\text{F}^2}{2\pi}\left(\partial_{\xpsr} \DM(\xpsr)\right)^2
\end{align}
where we used the lens equation (\ref{eq:lenseq}), and use that $x_\text{psr}\simeq x$ twice in the second line for the Taylor expansion and the change of variables. 
We take the second term on the right-hand side as the model error on the DM, which is propagated to the magnifications as described later in this section.

We first convert $\DM(t)$ to a spatial function of the lens plane $x$, taking the lens plane to be at the distance of the companion. 
This is done using an effective velocity, the only parameter of the lens:
\begin{equation}
\label{eq:effectivevelocity}
    \frac{x}{R_\text{F}} = \frac{t\,\vrel}{R_\text{F}\sqrt{d_\text{pl}/a_p \sin i}} \equiv \frac{t\,\veff}{R_\text{F}},
\end{equation}
where $\vrel = v_\text{psr} - v_\text{lens}$ is the binary relative velocity, $R_\text{F} = \sqrt{\lambda a_p\sin i}$ is a fiducial Fresnel scale, and $x$ is along the direction of $\vrel$ (See Fig.~\ref{fig:sub-second}). 
In this section, we set $\sqrt{d_\text{pl}/a_p\sin i}=1$, i.e. assume that lensing happens at binary separation projected on to the line of sight, and discuss the effect of changing $d_\text{pl}$ in Section~\ref{sec:physics}. 
From $\DM(x)$, we use the lens equation (\ref{eq:lenseq}), giving an explicit mapping from the lens plane to the source plane $\xpsr(x)$. 
We are, however, interested in the inverse map; for each source position $\xpsr$, we want to know where the images $x_i$ are on the lens plane. 

Because the pulse profile is integrated to 10\,s, the magnification in each time bin is the actual magnification averaged over the length of the time bin. 
The DM values are also limited to a 10\,s resolution, but it is actually necessary to compute magnifications on a finer resolution, followed by averaging, to simulate the effect of 10\,s subintegrations. 
Therefore, we have to interpolate $\xpsr(x)$ below this scale to compute image locations and their associated magnifications. 
While there are a number of ways to do this, cubic splines are well suited for our data, since our measurement errors are small ($\sim$ 1 per cent) and solutions, i.e. images $x_i$, can be efficiently found due to the polynomial nature of the interpolation. 
This interpolation effectively makes the lens smooth below the \SI{10}{s} time-scale. 

In addition to the `artificial' smoothing from our interpolation, the DM also appears to be physically smoothed. Interpreting the scattered images to be distributed in an extended disc effectively creates a frequency-dependent spatial filter on the observed DM since the DM must be sampled from all the scattered images (e.g. \citealt{cordes+16}).
The largest scattering times in our observation $\tau_\text{scat} \approx \SI{350}{\us}$ corresponds to a physical size of $\approx \SI{10 000}{km}$, corresponding to $\approx \SI{20}{s}$ at our best-fitting $v_\text{eff}$, larger than our data resolution. This is a fundamental smoothness that cannot be overcome by sampling at finer time resolutions. On the other hand, this justifies our artificial smoothing from interpolation. 
The combination of the relative closeness of these two smoothing scale is likely what makes our modelling so successful, despite the many shortcomings described in Section~\ref{sec:limitations}.

Once we have an interpolated function $\xpsr(x)$, we can solve the lens equation (\ref{eq:lenseq}) for a grid of $x_{\text{psr},j}$, finding images $x_i(x_{\text{psr},j})$, and associated image magnifications $\mu_i(x_{\text{psr},j})$ and total magnification from Equations (\ref{eq:magnification}) and (\ref{eq:mugeom}), respectively. 
The total magnification, rebinned to the resolution of the data, can be used to directly fit for $v_\text{eff}$ using a standard chi-square minimization routine. We use 90 data points from the frequency-averaged magnification between roughly orbital phase 0.187 and 0.293 for the fit.

The measured magnification, along with the best-fitting model, with $v_\text{eff} = \SI[separate-uncertainty=true]{470\pm 10}{km.s^{-1}}$, 
is shown in the bottom left-hand panel of Fig.~\ref{fig:fit}, while the plot of $\chi^{2}$ versus $v_{\rm eff}$ is shown in Fig.~\ref{fig:veff}. 
The error bars on the model are from the standard deviations of the magnification curve computed from 1000 Monte Carlo realizations of DM with 1$\sigma$ error added, where the error on DM is taken as the measurement error and bias term from equation (\ref{eq:dmerr}) added in quadrature. 
Note that the bias term equation (\ref{eq:dmerr}) depends on $\veff$, so we employ a two-step fitting: first, we minimize without the bias to get an approximate $\veff$, then we use the approximate $\veff$ to compute the bias term and fit a second time to get the proper errors. 
We also evaluate the magnification using the best-fitting model at all frequencies, and multiply the result with the $\mu_\text{ISS}$ to directly compare to the observed dynamic spectrum $I_\text{obs}(t, \nu)$, in the middle and top panels of Fig.~\ref{fig:fit}, respectively. 

Qualitatively, the locations of the `peaks and troughs' from the model closely match the data. 
The amplitude of variations in the model matches the data well, but are not a perfect one-to-one correlation, as evinced by the large $\chi^{2}$ value.  We emphasize again that this is a one-parameter fit, and we discuss some limitations of our models in Section~\ref{sec:limitations}.

As a cross-check, we use our best-fitting velocity $v_\text{eff} = \SI{470}{km.s^{-1}}$ to compute the magnifications in the second eclipse, where the ISS is at a minimum, shown on the right-hand panel of Fig.~\ref{fig:fit}. 
Though the DM is measured much less precisely in the second eclipse, the locations of the peaks and troughs in the simulated magnification curves, and the overall amplitude of fluctuation, also match the data well.

By changing $\lambda$, or equivalently $\nu$ for a given $v_\text{eff}$, the model gives a prediction of lensing for any frequency. 
We show the magnification spectrum for our best-fitting $v_\text{eff} = \SI{470}{km.s^{-1}}$, and at half and twice that value in the top panels, respectively, across over \text{a decade} of frequency: 200--4000 MHz, with the white-dashed lines showing our observational bandwidth. 
As the effective velocity is lowered, caustics, shown as the bright cusp-like features, become more prevalent at higher frequencies. 
Our model predicts the formation of many caustics at lower frequencies, and well-behaved weak-lensing modulations at high frequencies.  We briefly discuss future possibilities of wide-band observations in our conclusions.

\section{Interpretation of model velocity}
\label{sec:physics}
While we can only measure the DM as a function of time $\DM(t)$, our geometric model depends on the column density as a function of distance in the lens plane $\DM(x)$, in units of the Fresnel scale.  We can convert $\DM(t)$ into a spatial gradient using an effective velocity $v_\text{eff}$ and the distance of the lens, as in equation~(\ref{eq:effectivevelocity}).
The model fits explicitly constrain $\veff$.  Interpreting this effective velocity is not so straightforward, as it is dependent on the orbital properties of the system, the lens distance, and a flow velocity of the lensing material.

From timing \citep{stappersOrbitalEvolutionProper1998}, the projected semimajor axis $X = a_p \sin i = \SI{0.045 076 \pm 0.000 001}{lt\text{-}s}$ (denoted here with $X$ to avoid confusion with the $x$-axis), and the radial velocity amplitude $K_p = 2\pi a_p \sin i/P_b = \SI{9.9 156 \pm 0.0002}{km.s^{-1}}$ are measured. We express the following equations in terms of these observables.
The orbital separation is
\begin{align}
    a_\text{orb} = \frac{X}{\sin i} \frac{m_\text{p}+m_\text{c}}{m_\text{c}},
\end{align}
and the relative orbital velocity is
\begin{align}
    v_\text{rel} = \frac{2\pi a_\text{orb}}{P_b}  = \frac{K_p}{\sin i} \frac{m_\text{p}+m_\text{c}}{m_\text{c}},
\end{align}
where $m_\text{p}$ and $m_\text{c}$ are the masses of the pulsar and the companions, respectively.

The parameter $\veff$ is constrained by our geometric model to be
\begin{equation}
    \veff = \frac{1}{\sqrt{d_\text{pl}/X}}\left(v_\text{rel} - v_{\rm flow,\perp}\right),
\end{equation}
with the remaining unknowns being the flow velocity $v_\text{flow}$ projected onto the axis of motion and the ratio of lens distance to the orbital separation $d_\text{pl}/X$ projected onto the line of sight.
In \psr,  there are too many unknowns to achieve a meaningful constraint, as the mass ratio and $i$ are poorly constrained.  We can however, test with plausible values. Using $i \sim \SI{40}{deg}$ from \citet{stappersLightCurveCompanion1998}, assuming $m_\text{p} / m_\text{c} \sim 36$ as \citet{lazaridisEvidenceGravitationalQuadrupole2011} and setting $d_\text{pl}=X$ gives $\veff \approx \SI{570}{km.s^{-1}}$ .  

Our best-fitting value for $\veff = \SI[separate-uncertainty=true]{470\pm 10}{km.s^{-1}}$ differs from the above estimate by \SI{100}{km.s^{-1}}, and could be easily explained by a higher inclination of $i\approx \SI{51}{deg}$, a lower mass ratio of $m_\text{p} / m_\text{c} \approx 30$, or the lensing occurring at a distance of $d_\text{pl}\approx 1.5 X$.
Our constraint is then consistent with the lensing material co-moving with the companion; the plasma contributing to the excess DM surrounding eclipse does not have a transverse flow velocity greater than 100s of \si{km.s^{-1}}. 
\citet{polzin+19} estimates an outflow velocity of order \SI{5000}{km.s^{-1}} from momentum flux balance of the ablated material from the companion and the pulsar wind at orbital separation. 
If such a large outflow velocity is present, then it is predominantly along the line of sight, i.e. away from the pulsar.

\begin{figure}
    \begin{center}
        \begin{adjustbox}{clip,trim=0.8cm 0.0cm 0.70cm 0.6cm}
        \input{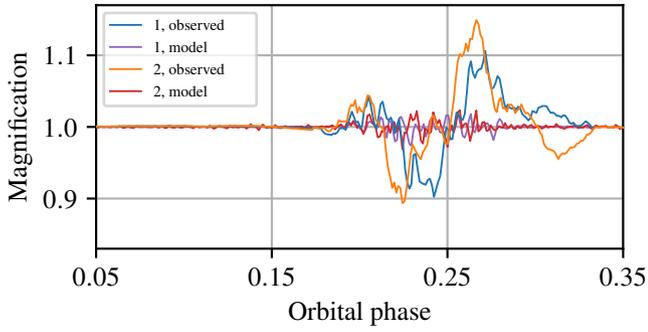}
        \end{adjustbox}
    \end{center}
    \vspace{-5mm}
    \caption{
    Measured and model magnification curves for the two eclipses with a 3 min moving average window. Blue and purple denote the observed and model curves for the first eclipse, and orange and red for the second, respectively. Both measured curves show a similar morphology, and this broad-scale effect is not captured by our lensing model.
    }
    \label{fig:mag_rebinned}
\end{figure}

We have presently modelled the effective velocity as a constant, and changes of velocity and distances over the eclipse are not captured by the model. 
For example, using the geometry in Fig.~\ref{fig:2Dmodel}, modelling the orbit as circular and the eclipsing plasma as a spherical cloud, the distance can vary by over 5 per cent over the eclipse. This will induce a difference in lensing which is symmetric about the orbital phase of 0.25. The change in velocity, on the other hand, will induce an antisymmetric difference. 
A combination of these changes may explain the biases seen in Fig.~\ref{fig:mag_rebinned}.
These effects introduce additional freedom in the model, and were not explored in this paper.

\section{Limitations / extensions of our model}
\label{sec:limitations}
As shown in Fig.~\ref{fig:fit},
the model fits observed magnification well for 1D, single parameter model.
However, our model is known to be incomplete, as can be seen by some unmodelled effects in our data. 
For example, Fig.~\ref{fig:mag_rebinned} shows a broad-scale feature in the magnification throughout both eclipses at the 10 per cent level, which is not captured in our model.

In this section, we discuss some simplifying assumptions in the preceding discussion and list some potential limitations of our incomplete model. 
These limitations are a non-exhaustive list of effects to 
consider in future work,
whether for eclipse lensing, or for lensing in the interstellar medium (ISM).

\subsection{Single image versus scattered Images}
\label{sec:limitations_geomoptics}

Our best-fitting $\veff$ puts our observational band in the one-image regime. Interestingly, the single image regime also seems to be at odds with the measured excess scattering in the eclipse.
Scattering implies that the screen produces many images, which are explicitly not captured in the model. 
The fact that the model works at all is thus rather surprising. 
However, the DM is well constrained, with only about 1 per cent relative error during the eclipse. 
If there are many images produced due to smaller scale fluctuations in the DM at the 1 per cent level, it may be that the magnification of those images average out to the one-image value.

As noted in equation~(\ref{eq:groupdelay}), the pulsar emission refracted by DM throughout the eclipse also has a group delay from the combined geometric and dispersive delays. 
We can do a simple estimate of the contributions from the maximal observed gradient of DM to check whether it is consistent with the scattering time.
The dispersive term's contribution is on the order of
\begin{align}
    \tau_{\text{group, disp},\,i} \approx \SI{41}{\us} \left(\frac{\lambda}{\SI{21}{cm}}\right)^2\left(\frac{\DM(x)|_{x=x_i}}{\SI{0.02}{pc.cm^{-3}}}\right). 
\end{align}
For the geometric delay, substituting equation~(\ref{eq:lenseq}) into the geometric term of equation~(\ref{eq:groupdelay}), we get
\begin{align}
    \tau_{\text{group, geom},\,i} \approx \SI{6}{\us} \left(\frac{\lambda}{\SI{21}{cm}}\right)^4\left(\frac{\Delta \DM\Big/\Delta x}{\SI{0.02}{pc.cm^{-3}}\Big/4700\,\text{km}}\right)^2,
\end{align}
where we used the best-fitting value of $v_\text{eff} = \SI{470}{km.s^{-1}}$ over \SI{10}{s} to estimate the gradient, and assumed that lensing happens at the binary separation for $R_\text{F}$. 
Note that, in general, the wavelength scaling of the geometric delay does not scale as $\sim \lambda^4$, but depends on how exactly images move in the model in time and frequency.
However, in our case, it is $\sim$ 1--2 order of magnitudes smaller than the observed scattering time, thus the variable scattering observed in the eclipse cannot be due to the geometric delay of the refracted image. 

A related effect is that, as discussed in Section~\ref{sec:magfromdm}, the scattering disc size is $\approx \SI{10 000}{km}$ at binary separation, which corresponds to $\approx \SI{20}{s}$ at our best-fitting velocity. 
This means that if the measured DM comes from the combination of all the images in the scattering disc, we are effectively measuring the physical column density smoothed by a spatial Gaussian filter. We might, in fact, be measuring the magnification of a 
disc of scattered images which is being refracted by the observed large scale DM fluctuations. Conversely, assuming the scattering occurs at the same distance as lensing, the scattering disc size lets us infer a resolution of the lens. 
We discuss the implication of this in Section~\ref{sec:eclipsescreen}.

Random phases of the scattered micro-images essentially broadens the image, thus tempering any caustic divergences.
However, caustics have been seen to occur in PSR~B1957\,$+$\,20 in a region where the fluctuations in dispersive phase closely match that of the geometric phase \citep{main+18}.
If caustics exist in \psr, they would only be visible in the edge of the eclipse, where there is still excess DM, but scattering is minimal.

\subsection{Screen geometry}
\label{sec:limitations_screendim}
\begin{figure}
    \begin{center}
        \begin{adjustbox}{clip,trim=0.8cm 0.0cm 0.7cm 0.6cm}
        \input{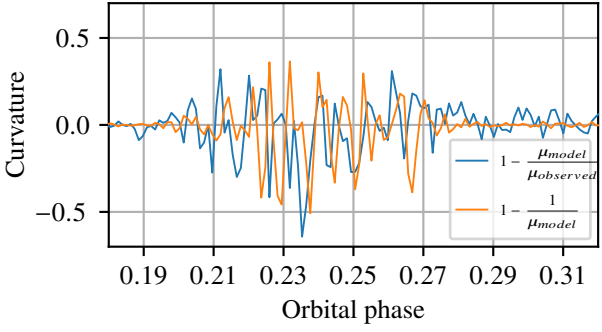}
        \end{adjustbox}
    \end{center}
        \vspace{-5mm}

    \caption{
    Lens curvatures from reciprocal magnifications from equation~(\ref{eq:ycurvature}) and (\ref{eq:magnification}). See Section~\ref{sec:limitations_screendim} for a description.
    }
    \label{fig:ycurvature}
\end{figure}

The physical column density can vary in both dimensions transverse to the line of sight, but the pulsar's motion only samples the column density along a single line in the plane. 
In our model, we assumed that the lens is 1D, and all lensing happens along the direction of the source motion. 
Although a perfect 1D lens seems unrealistic, it is not a simple task to measure or model the screen in the transverse dimension. 
In a 1D lens, light can only be bent in the said dimension, and thus, by flux conservation, magnification averages to 1 over a length-scale comparable to the lens size. 
However, fluctuations of the lens in the transverse dimension can focus or defocus light into or out of the line of source motion, causing a bias in modelled magnification.
We can estimate the contribution of the transverse direction, assuming that the unmodelled magnification is due to it. Indeed, if the physical column density represented by the measured $\DM$ depended on both dimensions $(x,y)$, the resulting magnification would be
\begin{align}
    \label{eq:mag2d}
    \mu_\text{2D} = \bigg(1&-\frac{\lambda r_\text{e} R_\text{F}^2}{2\pi}(\partial_x^2 \DM+\partial_y^2 \DM) + \nonumber\\ &\left(\frac{\lambda r_\text{e} R_\text{F}^2}{2\pi}\right)^2\left(\partial_x^2 \DM\,\partial_y^2 \DM-(\partial_x\partial_y \DM)^2\right)\bigg)^{-1}\bigg.
\end{align}
Combining this with the 1D magnification from equation~(\ref{eq:magnification}), we get, in the single image regime,
\begin{align}
    \label{eq:ycurvature}
    1-\frac{\mu_\text{1D}}{\mu_\text{2D}} = \frac{\lambda r_\text{e} R_\text{F}^2}{2\pi} \partial_y^2 \DM + \mu_\text{1D} \left(\frac{\lambda r_\text{e} R_\text{F}^2}{2\pi}\partial_x\partial_y \DM\right)^2,
\end{align}
where the RHS is due to deflection in the transverse direction. Taking $\mu_\text{1D}=\mu_\text{model}$ and $\mu_\text{2D}=\mu_\text{ecl}$ the measured magnification and interpreting the unmodeled magnification as solely coming from the transverse direction gives the contribution of the transverse direction to magnification. If we further assume that the cross derivative term is small, i.e. that the lenses are roughly isotropic, we can estimate curvature in the column density in the transverse direction.  Fig.~\ref{fig:ycurvature} shows the dimensionless `curvature' in equation~(\ref{eq:ycurvature}) compared to the the 1D contribution $1-1/\mu_\text{1D}=\frac{\lambda r_\text{e} R_\text{F}^2}{2\pi} \partial_x^2 \DM$, which appear to be of similar order, suggesting that the differences between our model and the observed magnifications can be explained through variations of geometry between the lenses.

In addition to the effect of eclipse geometry discussed in the end of Section~\ref{sec:physics}, a transverse dimension may also contribute to the systematic deviations of order $\sim 0.1$ from magnification of unity in 3-min rolling-averaged magnification curves of both eclipses in Fig.~\ref{fig:mag_rebinned}. 
An overdensity of electrons in the transverse dimension in orbital phase $\lesssim 0.25$ followed by an underdensity in orbital phase $\gtrsim 0.25$ can lead to the observed bias. 

In principle, there can be an extra angle in the effective velocity, describing the relative angle between the orientation of the 1D lenses and the motion of the pulsar on the sky, similar to the angle $\alpha$ in pulsar scintillation \citep[e.g.][]{brisken+10}. 
However, unlike in the context of pulsar scintillation, a constant angle is physically unlikely. 
The orientation of the lenses would more likely either follow the contours of Fig.~\ref{fig:sub-second}, and thus slowly change over the orbit, or be a statistical average over many angles from the many anisotropic lensing plasma `blobs'. 
Modelling such lenses requires a full 2D simulation, beyond the scope of this paper.

\subsection{Effects of screen resolution}
\label{sec:eclipsescreen}
We have analysed the effects of flux density variations from lensing assuming a point source. However, the large scattering times from the eclipsing wind result in extremely fine resolution at the pulsar. 
Similarly, the flux density variations we observe are the result of a two-screen interaction, where 
we have treated the lensing from the eclipsing material as uncoupled from scintillation in the ISM.
For a screen, the angular size relates to the time delay as 
\begin{align}
    \label{eq:angledelay}
    \tau = \frac{\theta^{2}}{2c}\frac{d_\text{psr}\,d_\text{lens}}{d_\text{psr} - d_\text{lens}},
\end{align}
where $d_\text{psr}$, $d_\text{lens}$, and $c$ are the distances to the pulsar, to the lens, and the speed of light, respectively. The screen resolution given by the diffractive limit of the screen is then
\begin{align}
    \label{eq:angularresolution}
    \theta_\text{res} = \frac{\lambda}{\sqrt{2c\tau}} \sqrt{\frac{d_\text{psr}/d_\text{lens}}{d_\text{psr}-d_\text{lens}}}.
\end{align}

In the first eclipse, the measured scattering times in the eclipsing region is as large as $\tau_\text{scat}\approx \SI{350}{\us}$. 
So the physical resolution is $x_\text{res}=d_\text{pl}\theta_\text{res}\approx \SI{10}{m}$,
much smaller than the neutron star radius of $R_{\rm NS} \approx \SI{10}{km}$, and far smaller still than the light cylinder radius, of $R_{\rm LC} \approx \SI{215}{km}$. Therefore, it is quite plausible that the emission region is resolved. 
An extended source generally has the effect of smoothing out magnification curves over the source size \citep[e.g.][]{cleggGaussianPlasmaLens1998}.
However, we are modelling magnification on \SI{10}{s} time bins, or equivalently, over \SI{4700}{km} at our best-fitting $\veff$, much larger than potential source-size effects, and any smoothing effects by an extended source will therefore be unlikely to influence the current model. 
The apparent lack of caustics may be also explained by the scattered images, as discussed in Section~\ref{sec:limitations_geomoptics}.

For the ISM and eclipse-screen interaction, if the pulsar's emission is sufficiently broadened before it reaches the ISM, it will no longer appear `point-like' to the interstellar screen, and will no longer scintillate.  Equivalently, emission from the many scattered images will be out of phase with each other, and it will appear as an incoherent source. 
This argument has been invoked in FRBs \citep{masui+15}, where scattering occurs both in the host and in our galaxy, and in the Crab Pulsar \citep{vandenberg76, rudnitskii+17, main+18b}, where scattering in the Nebula dominates the temporal broadening while the angular broadening is in the ISM. 

While we do not know the angular scale or distance to the scattering screen, we can approximate it as halfway, and use the scintillation bandwidth to approximate the screen's resolution.  
Although there is a wide-band scintle covering our entire bandwidth, there is clear variation from scintillation from channel to channel, with channel sizes of 1.56\,MHz. 
This channel to channel variation comes from the furthest separated images on the sky, and hence have the highest resolution. 
This implies scattered power to $\tau_\text{ISS} \sim 1 / (2\times \SI{1.56}{MHz}) \approx \SI{320}{ns}$.  

The condition for the ISM screen to resolve the eclipsing scattering screen is then roughly
\begin{equation}
    \frac{\sqrt{2c \tau_\text{scat}  d_\text{pl}}}{d_{\rm psr}/2}  \gtrsim \lambda\frac{\sqrt{2}}{\sqrt{d_{\rm psr}c\tau_\text{ISS}}}.
\end{equation}
From the estimated values, the size of scattering at the eclipse is $\theta_{\rm ecl} \approx 90$\,nas, while the resolution of the ISM scattering screen is $\theta_{\rm res} \approx 900$\,nas, an order of magnitude larger than the scattering in the eclipse.  It is then unlikely that this will affect our observation, although the largest time delays (finest scintles) of the ISM screen may be at the limit of resolving the largest time delays of the eclipsing screen. 

While this is unlikely to affect our observation, it is very likely to matter at lower frequencies; using $\tau \propto \lambda^{4}$, the angular size and resolution scale as $\theta_\text{ecl} / \theta_\text{res} \propto \lambda^{3}$.
In that case, during the eclipse, the maxima and minima of the scintles in the dynamic spectrum will tend to be more `washed out'.

\section{Conclusions and further work}
\label{sec:summary}

In this paper, we discovered clear evidence of lensing associated with the eclipsing material of \psr, and demonstrated a direct, predictive correspondence between DM and flux density variations in the pulsar, using a simple 1D lensing model. 
This provided a measure of the effective velocity of the lensing material, constraining the flow velocity to be small, and thus co-moving with the companion; or predominantly along the line of sight, away from the pulsar.
A single measurement of an effective velocity is difficult to interpret due to the unknown orbital parameters and degeneracies.  Differential measurements are far more precise -- if $\veff$ surrounding the eclipse changes over time, then this can only arise from variations in the companion's outflow, as the proper motion and orbital dynamics are essentially constant. The more that is known about the system (e.g. mass ratio, inclination, precise pulsar distance), the better this constraint can be used to probe the outflow, and possibly eclipse dynamics. The approach taken in this paper may be even better applied to other spider systems, with better understood mass ratios and orbits, if measurements are sufficiently precise to allow for it.

Our predictive framework may be useful when applied to the ISM, i.e. if scattering and flux density variations caused by variable DM in scattering screens can be predicted.  This will be especially the case in systems with solved scattering screen geometry, either through VLBI or multistation time delays \citep{brisken+10, fadeev+18, simard+19b}, or through annual variations of scintillation time-scales and bandwidths \citep{rickett+14, reardon+19} or scintillation arc curvatures \citep{mainMeasuringInterstellarDelays2020,reardonPrecisionOrbitalDynamics2020}.  

In addition to predicting the the flux density variations, the model also predicts the associated scattering times of many images.
Unfortunately, as shown in Section~\ref{sec:eclipsescreen}, predicting scattering times was unsuccessful in our present model; scattering is highly dependent on the smallest scales of fluctuations in DM, which are most difficult to measure precisely.
However, with increasingly precise measures of DM, it may be possible to use DM variations to predict or estimate the effects of variable scattering in the ISM to improve timing.
This type of analysis may already be possible using existing high-cadence DM monitoring campaigns, including Pulsar Timing Arrays (\citealt{Hobbs2013}, \citealt{desvignes+16}, \citealt{verbiest+16}, \citealt{arzoumanian+18}), and high-precision DM monitoring at low frequencies \citep[e.g.][]{donner+19,bondonneauPulsarsNenuFARBackend2020}. 
As the CHIME Pulsar Project operates for longer \citep{CHIMEpulsar}, there will be daily DM measurements for most pulsars in the northern sky, a perfect test-bed for the causal connection between the variations of DM, flux density, and scattering.

One of the limitations of this paper comes from the \SI{10}{s} time binning, as there is clearly lensing occurring on shorter time-scales, and perhaps caustics as seen in PSR~B1957\,$+$\,20 \citep{main+18} and Ter5A \citep{bilous+19}. In \psr, \citet{polzin+19} find depolarization over 13\,min bins throughout the eclipsing region, indicating rapid RM changes.  Further studies able to resolve single pulses will be useful to detect caustics, and are much more likely to provide instantaneous measures of the RM of the eclipsing material \citep{li+19}, and potentially resolve the pulsar magnetosphere (e.g. \citealt{main+18}). 
On the other hand, it may be difficult to accurately measure DM at much shorter time-scales. In addition, DM measurements at lower frequencies are biased by larger bending angles as well as the presence of multiple images, and if there is significant scattering, the DM is effectively smoothed by the scattering discs \citep{cordes+16}. Therefore, higher frequency measurements, which are less affected by various propagation effects, may be more suitable for the present type of predictive modelling.
Further studies using sensitive, wide-band observations (such as those afforded by e.g.  MeerKAT, LOFAR, FAST, Parkes UWL), will be necessary to further unravel the mystery of eclipses.

\section*{Acknowledgements}

We are indebted to Stefan Os\l{}owski, whose previous work on this dataset motivated our current work. 
We thank Marten van Kerkwijk for discussions and for contributing many helpful comments on an early draft of the paper. We also greatly appreciate discussions and comments from Daniel Baker, Ue-Li Pen, and Olaf Wucknitz, as well as detailed comments from the anonymous referee. 
FXL is extremely grateful to Ue-Li Pen for supervision and support, FXL thanks the Max-Planck-Institut f\"{u}r Radioastronomie, where part of this work was done, and the University of Toronto Scintillometry group for providing helpful feedback throughout the project. 
JPWV acknowledges support by the Deutsche Forschungsgemeinschaft (DFG)
through the Heisenberg programme (Project No. 433075039).
GS acknowledges support
from the Netherlands Organisation for Scientific Research
NWO (TOP2.614.001.602).

This work was based on observations with the 100-m radio telescope of the MPIfR (Max-Planck-Institut f{\"u}r Radioastronomie) at Effelsberg, Germany. 
The analysis exntensively used the \textsc{scipy}, \textsc{numpy} \citep{2020SciPy-NMeth}, and \textsc{astropy} \citep{astropy:2013,astropy:2018} packages.

\section*{Data availability}
The data underlying this article is available upon request to the corresponding author.

\bibliographystyle{mnras}
\bibliography{J2051}

\begin{thebibliography}{}
\makeatletter
\relax
\def\mn@urlcharsother{\let\do\@makeother \do\$\do\&\do\#\do\^\do\_\do\%\do\~}
\def\mn@doi{\begingroup\mn@urlcharsother \@ifnextchar [ {\mn@doi@}
  {\mn@doi@[]}}
\def\mn@doi@[#1]#2{\def\@tempa{#1}\ifx\@tempa\@empty \href
  {http://dx.doi.org/#2} {doi:#2}\else \href {http://dx.doi.org/#2} {#1}\fi
  \endgroup}
\def\mn@eprint#1#2{\mn@eprint@#1:#2::\@nil}
\def\mn@eprint@arXiv#1{\href {http://arxiv.org/abs/#1} {{\tt arXiv:#1}}}
\def\mn@eprint@dblp#1{\href {http://dblp.uni-trier.de/rec/bibtex/#1.xml}
  {dblp:#1}}
\def\mn@eprint@#1:#2:#3:#4\@nil{\def\@tempa {#1}\def\@tempb {#2}\def\@tempc
  {#3}\ifx \@tempc \@empty \let \@tempc \@tempb \let \@tempb \@tempa \fi \ifx
  \@tempb \@empty \def\@tempb {arXiv}\fi \@ifundefined
  {mn@eprint@\@tempb}{\@tempb:\@tempc}{\expandafter \expandafter \csname
  mn@eprint@\@tempb\endcsname \expandafter{\@tempc}}}

\bibitem[\protect\citeauthoryear{Arzoumanian et~al.,}{Arzoumanian
  et~al.}{2018}]{arzoumanian+18}
Arzoumanian Z.,  et~al., 2018, \mn@doi [\apjs] {10.3847/1538-4365/aab5b0}, 235,
  37

\bibitem[\protect\citeauthoryear{{Astropy Collaboration}}{{Astropy
  Collaboration}}{2013}]{astropy:2013}
{Astropy Collaboration} 2013, \mn@doi [\aap] {10.1051/0004-6361/201322068},
  \href {http://adsabs.harvard.edu/abs/2013A%26A...558A..33A} {558, A33}

\bibitem[\protect\citeauthoryear{{Astropy Collaboration}}{{Astropy
  Collaboration}}{2018}]{astropy:2018}
{Astropy Collaboration} 2018, \mn@doi [AJ] {10.3847/1538-3881/aabc4f}, \href
  {https://ui.adsabs.harvard.edu/abs/2018AJ....156..123A} {156, 123}

\bibitem[\protect\citeauthoryear{{Backer}, {Kulkarni}, {Heiles}, {Davis}  \&
  {Goss}}{{Backer} et~al.}{1982}]{backer+82}
{Backer} D.~C.,  {Kulkarni} S.~R.,  {Heiles} C.,  {Davis} M.~M.,   {Goss}
  W.~M.,  1982, \mn@doi [\nat] {10.1038/300615a0}, \href
  {http://adsabs.harvard.edu/abs/1982Natur.300..615B} {300, 615}

\bibitem[\protect\citeauthoryear{{Bilous}, {Ransom}  \& {Demorest}}{{Bilous}
  et~al.}{2019}]{bilous+19}
{Bilous} A.~V.,  {Ransom} S.~M.,   {Demorest} P.,  2019, \mn@doi [\apj]
  {10.3847/1538-4357/ab16dd}, \href
  {https://ui.adsabs.harvard.edu/abs/2019ApJ...877..125B} {877, 125}

\bibitem[\protect\citeauthoryear{Bondonneau et~al.,}{Bondonneau
  et~al.}{2020}]{bondonneauPulsarsNenuFARBackend2020}
Bondonneau L.,  et~al., 2020, arXiv e-prints, 2009, arXiv:2009.02076

\bibitem[\protect\citeauthoryear{{Brisken}, {Macquart}, {Gao}, {Rickett},
  {Coles}, {Deller}, {Tingay}  \& {West}}{{Brisken} et~al.}{2010}]{brisken+10}
{Brisken} W.~F.,  {Macquart} J.~P.,  {Gao} J.~J.,  {Rickett} B.~J.,  {Coles}
  W.~A.,  {Deller} A.~T.,  {Tingay} S.~J.,   {West} C.~J.,  2010, \mn@doi
  [\apj] {10.1088/0004-637X/708/1/232}, \href
  {https://ui.adsabs.harvard.edu/abs/2010ApJ...708..232B} {708, 232}

\bibitem[\protect\citeauthoryear{Clegg, Fey  \& Lazio}{Clegg
  et~al.}{1998}]{cleggGaussianPlasmaLens1998}
Clegg A.~W.,  Fey A.~L.,   Lazio T. J.~W.,  1998, \mn@doi [\apj] {10/bfxdkz},
  496, 253

\bibitem[\protect\citeauthoryear{{Cordes}, {Shannon}  \& {Stinebring}}{{Cordes}
  et~al.}{2016}]{cordes+16}
{Cordes} J.~M.,  {Shannon} R.~M.,   {Stinebring} D.~R.,  2016, \mn@doi [\apj]
  {10.3847/0004-637X/817/1/16}, \href
  {https://ui.adsabs.harvard.edu/abs/2016ApJ...817...16C} {817, 16}

\bibitem[\protect\citeauthoryear{Cordes, Wasserman, Hessels, Lazio, Chatterjee
  \& Wharton}{Cordes et~al.}{2017}]{cordesLensingFastRadio2017}
Cordes J.~M.,  Wasserman I.,  Hessels J. W.~T.,  Lazio T. J.~W.,  Chatterjee
  S.,   Wharton R.~S.,  2017, \mn@doi [\apj] {10/gf6nsn}, 842, 35

\bibitem[\protect\citeauthoryear{{Desvignes} et~al.,}{{Desvignes}
  et~al.}{2016}]{desvignes+16}
{Desvignes} G.,  et~al., 2016, \mn@doi [\mnras] {10.1093/mnras/stw483}, \href
  {https://ui.adsabs.harvard.edu/abs/2016MNRAS.458.3341D} {458, 3341}

\bibitem[\protect\citeauthoryear{{Donner} et~al.,}{{Donner}
  et~al.}{2019}]{donner+19}
{Donner} J.~Y.,  et~al., 2019, \mn@doi [\aap] {10.1051/0004-6361/201834059},
  \href {https://ui.adsabs.harvard.edu/abs/2019A&A...624A..22D} {624, A22}

\bibitem[\protect\citeauthoryear{{Fadeev}, {Andrianov}, {Burgin}, {Popov},
  {Rudnitskiy}, {Shishov}, {Smirnova}  \& {Zuga}}{{Fadeev}
  et~al.}{2018}]{fadeev+18}
{Fadeev} E.~N.,  {Andrianov} A.~S.,  {Burgin} M.~S.,  {Popov} M.~V.,
  {Rudnitskiy} A.~G.,  {Shishov} V.~I.,  {Smirnova} T.~V.,   {Zuga} V.~A.,
  2018, \mn@doi [\mnras] {10.1093/mnras/sty2055}, \href
  {https://ui.adsabs.harvard.edu/abs/2018MNRAS.480.4199F} {480, 4199}

\bibitem[\protect\citeauthoryear{{Fruchter} \& {Goss}}{{Fruchter} \&
  {Goss}}{1992}]{fruchter+92}
{Fruchter} A.~S.,  {Goss} W.~M.,  1992, \mn@doi [\apjl] {10.1086/186259}, \href
  {https://ui.adsabs.harvard.edu/abs/1992ApJ...384L..47F} {384, L47}

\bibitem[\protect\citeauthoryear{{Fruchter}, {Stinebring}  \&
  {Taylor}}{{Fruchter} et~al.}{1988}]{fruchter+88}
{Fruchter} A.~S.,  {Stinebring} D.~R.,   {Taylor} J.~H.,  1988, \mn@doi [\nat]
  {10.1038/333237a0}, \href {http://adsabs.harvard.edu/abs/1988Natur.333..237F}
  {333, 237}

\bibitem[\protect\citeauthoryear{{Hobbs}}{{Hobbs}}{2013}]{Hobbs2013}
{Hobbs} G.,  2013, \mn@doi [Classical and Quantum Gravity]
  {10.1088/0264-9381/30/22/224007}, \href
  {https://ui.adsabs.harvard.edu/abs/2013CQGra..30v4007H} {30, 224007}

\bibitem[\protect\citeauthoryear{{Kaspi}}{{Kaspi}}{2010}]{kaspi10}
{Kaspi} V.~M.,  2010, \mn@doi [Proceedings of the National Academy of Science]
  {10.1073/pnas.1000812107}, \href
  {http://adsabs.harvard.edu/abs/2010PNAS..107.7147K} {107, 7147}

\bibitem[\protect\citeauthoryear{Kulkarni}{Kulkarni}{2020}]{kulkarniDispersionMeasureConfusion2020}
Kulkarni S.~R.,  2020, arXiv e-prints, 2007, arXiv:2007.02886

\bibitem[\protect\citeauthoryear{Lazaridis et~al.,}{Lazaridis
  et~al.}{2011}]{lazaridisEvidenceGravitationalQuadrupole2011}
Lazaridis K.,  et~al., 2011, \mn@doi [\mnras]
  {10.1111/j.1365-2966.2011.18610.x}, 414, 3134

\bibitem[\protect\citeauthoryear{{Lazarus}, {Karuppusamy}, {Graikou},
  {Caballero}, {Champion}, {Lee}, {Verbiest}  \& {Kramer}}{{Lazarus}
  et~al.}{2016}]{lazarus+16}
{Lazarus} P.,  {Karuppusamy} R.,  {Graikou} E.,  {Caballero} R.~N.,  {Champion}
  D.~J.,  {Lee} K.~J.,  {Verbiest} J.~P.~W.,   {Kramer} M.,  2016, \mn@doi
  [\mnras] {10.1093/mnras/stw189}, \href
  {http://adsabs.harvard.edu/abs/2016MNRAS.458..868L} {458, 868}

\bibitem[\protect\citeauthoryear{{Li}, {Lin}, {Main}, {Pen}, {van Kerkwijk}  \&
  {Yang}}{{Li} et~al.}{2019}]{li+19}
{Li} D.,  {Lin} F.~X.,  {Main} R.,  {Pen} U.-L.,  {van Kerkwijk} M.~H.,
  {Yang} I.~S.,  2019, \mn@doi [\mnras] {10.1093/mnras/stz374}, \href
  {https://ui.adsabs.harvard.edu/abs/2019MNRAS.484.5723L} {484, 5723}

\bibitem[\protect\citeauthoryear{Main \& van Kerkwijk}{Main \& van
  Kerkwijk}{2017}]{main+18b}
Main R.,  van Kerkwijk M.~H.,  2017. {Camb. Univ. Press}, pp 83--83,
  \mn@doi{10.1017/S1743921317009176}

\bibitem[\protect\citeauthoryear{{Main} et~al.,}{{Main} et~al.}{2018}]{main+18}
{Main} R.,  et~al., 2018, \mn@doi [\nat] {10.1038/s41586-018-0133-z}, \href
  {https://ui.adsabs.harvard.edu/abs/2018Natur.557..522M} {557, 522}

\bibitem[\protect\citeauthoryear{Main et~al.,}{Main
  et~al.}{2020}]{mainMeasuringInterstellarDelays2020}
Main R.~A.,  et~al., 2020, \mn@doi [Monthly Notices of the Royal Astronomical
  Society] {10/gk7txq}, 499, 1468

\bibitem[\protect\citeauthoryear{{Manchester}, {Hobbs}, {Teoh}  \&
  {Hobbs}}{{Manchester} et~al.}{2005}]{manchester+05}
{Manchester} R.~N.,  {Hobbs} G.~B.,  {Teoh} A.,   {Hobbs} M.,  2005, \mn@doi
  [\aj] {10.1086/428488}, \href
  {http://adsabs.harvard.edu/abs/2005AJ....129.1993M} {129, 1993}

\bibitem[\protect\citeauthoryear{{Masui} et~al.,}{{Masui}
  et~al.}{2015}]{masui+15}
{Masui} K.,  et~al., 2015, \mn@doi [\nat] {10.1038/nature15769}, \href
  {https://ui.adsabs.harvard.edu/abs/2015Natur.528..523M} {528, 523}

\bibitem[\protect\citeauthoryear{Ng}{Ng}{2017}]{CHIMEpulsar}
Ng C.,  2017. Camb. Univ. Press, pp 179--182,
  \mn@doi{10.1017/S1743921317010638}

\bibitem[\protect\citeauthoryear{Pennucci, Demorest  \& Ransom}{Pennucci
  et~al.}{2014}]{pennucciElementaryWidebandTiming2014}
Pennucci T.~T.,  Demorest P.~B.,   Ransom S.~M.,  2014, \mn@doi [ApJ]
  {10/ggpbmp}, 790, 93

\bibitem[\protect\citeauthoryear{{Planck Collaboration}}{{Planck
  Collaboration}}{2020}]{planckcollaborationPlanck2018Results2018}
{Planck Collaboration} 2020, \mn@doi [Astronomy and Astrophysics]
  {10.1051/0004-6361/201833886}, 641, A8

\bibitem[\protect\citeauthoryear{{Polzin}, {Breton}, {Stappers},
  {Bhattacharyya}, {Janssen}, {Os{\l}owski}, {Roberts}  \& {Sobey}}{{Polzin}
  et~al.}{2019}]{polzin+19}
{Polzin} E.~J.,  {Breton} R.~P.,  {Stappers} B.~W.,  {Bhattacharyya} B.,
  {Janssen} G.~H.,  {Os{\l}owski} S.,  {Roberts} M.~S.~E.,   {Sobey} C.,  2019,
  \mn@doi [\mnras] {10.1093/mnras/stz2579}, \href
  {https://ui.adsabs.harvard.edu/abs/2019MNRAS.490..889P} {490, 889}

\bibitem[\protect\citeauthoryear{{Polzin}, {Breton}, {Bhattacharyya},
  {Scholte}, {Sobey}  \& {Stappers}}{{Polzin} et~al.}{2020}]{polzin+20}
{Polzin} E.~J.,  {Breton} R.~P.,  {Bhattacharyya} B.,  {Scholte} D.,  {Sobey}
  C.,   {Stappers} B.~W.,  2020, \mn@doi [\mnras] {10.1093/mnras/staa596},
  \href {https://ui.adsabs.harvard.edu/abs/2020MNRAS.494.2948P} {494, 2948}

\bibitem[\protect\citeauthoryear{Press, Teukolsky, Vetterling  \&
  Flannery}{Press et~al.}{2007}]{pressNumericalRecipes3rd2007}
Press W.~H.,  Teukolsky S.~A.,  Vetterling W.~T.,   Flannery B.~P.,  2007,
  Numerical {{Recipes}} 3rd {{Edition}}: {{The Art}} of {{Scientific
  Computing}}.
{Camb. Univ. Press}

\bibitem[\protect\citeauthoryear{{Reardon}, {Coles}, {Hobbs}, {Ord}, {Kerr},
  {Bailes}, {Bhat}  \& {Venkatraman Krishnan}}{{Reardon}
  et~al.}{2019}]{reardon+19}
{Reardon} D.~J.,  {Coles} W.~A.,  {Hobbs} G.,  {Ord} S.,  {Kerr} M.,  {Bailes}
  M.,  {Bhat} N.~D.~R.,   {Venkatraman Krishnan} V.,  2019, \mn@doi [\mnras]
  {10.1093/mnras/stz643}, \href
  {https://ui.adsabs.harvard.edu/abs/2019MNRAS.485.4389R} {485, 4389}

\bibitem[\protect\citeauthoryear{Reardon et~al.,}{Reardon
  et~al.}{2020}]{reardonPrecisionOrbitalDynamics2020}
Reardon D.~J.,  et~al., 2020, \mn@doi [ApJ] {10/gk7txr}, 904, 104

\bibitem[\protect\citeauthoryear{{Rickett} et~al.,}{{Rickett}
  et~al.}{2014}]{rickett+14}
{Rickett} B.~J.,  et~al., 2014, \mn@doi [\apj] {10.1088/0004-637X/787/2/161},
  \href {https://ui.adsabs.harvard.edu/abs/2014ApJ...787..161R} {787, 161}

\bibitem[\protect\citeauthoryear{{Ruderman}, {Shaham}  \& {Tavani}}{{Ruderman}
  et~al.}{1989}]{ruderman+89}
{Ruderman} M.,  {Shaham} J.,   {Tavani} M.,  1989, \mn@doi [\apj]
  {10.1086/167029}, \href {http://adsabs.harvard.edu/abs/1989ApJ...336..507R}
  {336, 507}

\bibitem[\protect\citeauthoryear{{Rudnitskii}, {Popov}  \&
  {Soglasnov}}{{Rudnitskii} et~al.}{2017}]{rudnitskii+17}
{Rudnitskii} A.~G.,  {Popov} M.~V.,   {Soglasnov} V.~A.,  2017, \mn@doi
  [Astronomy Reports] {10.1134/S1063772917050043}, \href
  {https://ui.adsabs.harvard.edu/abs/2017ARep...61..393R} {61, 393}

\bibitem[\protect\citeauthoryear{{Shaifullah} et~al.,}{{Shaifullah}
  et~al.}{2016}]{shaifullah+16}
{Shaifullah} G.,  et~al., 2016, \mn@doi [\mnras] {10.1093/mnras/stw1737}, \href
  {http://adsabs.harvard.edu/abs/2016MNRAS.462.1029S} {462, 1029}

\bibitem[\protect\citeauthoryear{{Simard}, {Pen}, {Marthi}  \&
  {Brisken}}{{Simard} et~al.}{2019}]{simard+19b}
{Simard} D.,  {Pen} U.~L.,  {Marthi} V.~R.,   {Brisken} W.,  2019, \mn@doi
  [\mnras] {10.1093/mnras/stz2046}, \href
  {https://ui.adsabs.harvard.edu/abs/2019MNRAS.488.4963S} {488, 4963}

\bibitem[\protect\citeauthoryear{{Stappers} et~al.,}{{Stappers}
  et~al.}{1996}]{stappers+96}
{Stappers} B.~W.,  et~al., 1996, \mn@doi [\apjl] {10.1086/310148}, \href
  {http://adsabs.harvard.edu/abs/1996ApJ...465L.119S} {465, L119}

\bibitem[\protect\citeauthoryear{Stappers, Bailes, Manchester, Sandhu  \&
  Toscano}{Stappers et~al.}{1998a}]{stappersOrbitalEvolutionProper1998}
Stappers B.~W.,  Bailes M.,  Manchester R.~N.,  Sandhu J.~S.,   Toscano M.,
  1998a, \mn@doi [The Astrophysical Journal] {10/b7nsgd}, 499, L183

\bibitem[\protect\citeauthoryear{Stappers, van Kerkwijk, Lane  \&
  Kulkarni}{Stappers et~al.}{1998b}]{stappersLightCurveCompanion1998}
Stappers B.~W.,  van Kerkwijk M.~H.,  Lane B.,   Kulkarni S.~R.,  1998b,
  \mn@doi [The Astrophysical Journal Letters] {10.1086/311795}, 510, L45

\bibitem[\protect\citeauthoryear{Taylor}{Taylor}{1992}]{taylorPulsarTimingRelativistic1992}
Taylor J.~H.,  1992, \mn@doi [Philosophical Transactions of the Royal Society
  of London Series A] {10/fg4bh7}, 341, 117

\bibitem[\protect\citeauthoryear{{Thompson}, {Blandford}, {Evans}  \&
  {Phinney}}{{Thompson} et~al.}{1994}]{thompson+94}
{Thompson} C.,  {Blandford} R.~D.,  {Evans} C.~R.,   {Phinney} E.~S.,  1994,
  \mn@doi [\apj] {10.1086/173728}, \href
  {http://adsabs.harvard.edu/abs/1994ApJ...422..304T} {422, 304}

\bibitem[\protect\citeauthoryear{{Vandenberg}}{{Vandenberg}}{1976}]{vandenberg76}
{Vandenberg} N.~R.,  1976, \mn@doi [\apj] {10.1086/154753}, \href
  {https://ui.adsabs.harvard.edu/abs/1976ApJ...209..578V} {209, 578}

\bibitem[\protect\citeauthoryear{Verbiest et~al.,}{Verbiest
  et~al.}{2016}]{verbiest+16}
Verbiest J. P.~W.,  et~al., 2016, \mn@doi [\mnras] {10.1093/mnras/stw347}, 458,
  1267–1288

\bibitem[\protect\citeauthoryear{{Virtanen} et~al.,}{{Virtanen}
  et~al.}{2020}]{2020SciPy-NMeth}
{Virtanen} P.,  et~al., 2020, \mn@doi [Nature Methods]
  {https://doi.org/10.1038/s41592-019-0686-2}, \href {https://rdcu.be/b08Wh}
  {17, 261}

\bibitem[\protect\citeauthoryear{{van Straten} \& {Bailes}}{{van Straten} \&
  {Bailes}}{2011}]{vanStraten+11}
{van Straten} W.,  {Bailes} M.,  2011, \mn@doi [\pasa] {10.1071/AS10021}, \href
  {http://adsabs.harvard.edu/abs/2011PASA...28....1V} {28, 1}

\bibitem[\protect\citeauthoryear{{van Straten}, {Manchester}, {Johnston}  \&
  {Reynolds}}{{van Straten} et~al.}{2010}]{vanStraten+10}
{van Straten} W.,  {Manchester} R.~N.,  {Johnston} S.,   {Reynolds} J.~E.,
  2010, \mn@doi [\pasa] {10.1071/AS09084}, \href
  {http://adsabs.harvard.edu/abs/2010PASA...27..104V} {27, 104}

\makeatother
\end{thebibliography}

% \begin{onecolumn}
\begin{appendix}

\section{Inpainting dynamic spectra using a Wiener-filter}

\renewcommand{\thefigure}{A\arabic{figure}}
\begin{figure*}
    \begin{center}
        \begin{adjustbox}{clip,trim=0cm 0cm 0cm 1.5cm}
        \input{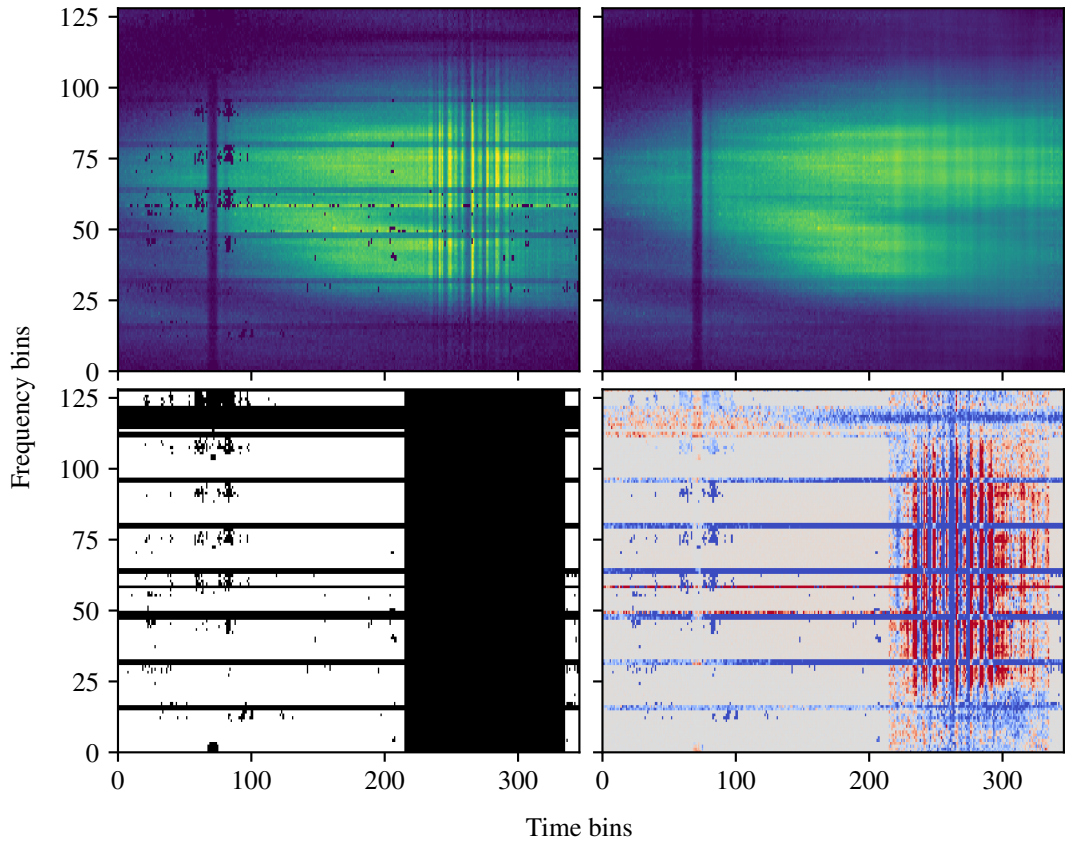}
        \end{adjustbox}
    \end{center}
    \vspace{-5mm}
    \caption{
    Wiener filter of the dynamic spectrum in the first eclipse. Top left: raw measured dynamic spectrum; top right: Wiener-filtered dynamic spectrum; bottom left: applied mask; bottom right: difference of raw and Wiener-filtered dynamic spectra.
    }
    \label{fig:wiener_dynspec}
\end{figure*}

Wiener filter is a linear filter often used in signal processing and image analyses, such as analyses of the Cosmic Microwave Background, to improve the SNR of data \citep[see e.g.][]{planckcollaborationPlanck2018Results2018}. 
Suppose we have a true signal vector $s$, e.g. a dynamic spectrum, or a time series. The measurement of $s$ may be corrupted, and the measured signal can be described by
\begin{align}
    d = H s + n
\end{align}
for some known matrix $H$ and additive noise vector $n$. $H$ can describe, for example, a boolean mask of the data, or a blurring process. 
The Wiener filter $W$ is the `optimal' filter that gives the estimated signal
\begin{align}
    s_\text{WF} = W d.
\end{align}
The Wiener filter is optimal in the least-squares sense: it minimizes the expected squared error
\begin{align}
    \varepsilon^2 = \langle|s_\text{WF} - s|^2\rangle.
\end{align}
Minimizing $\varepsilon^2$ with respect to $W$, and assuming that $s$ and $n$ are uncorrelated so that $\langle n s^\dagger\rangle= \langle s n^\dagger\rangle = 0$, the Wiener filter is
\begin{align}
    \label{eq:wienerfilter}
    W = \Sigma_s H^\dagger (H\Sigma_s H^\dagger + \Sigma_n)^{-1}
\end{align}
where $\dagger$ indicates the conjugate transpose, and $\Sigma_s = \langle s s^\dagger\rangle$ and $\Sigma_n = \langle n n^\dagger\rangle$ are the autocovariance matrices of the signal and noise, respectively. 
The diagonal of $\Sigma_s$ is the power in each pixel, and the off-diagonal term represents the covariance between pixels. 
Similarly, the diagonal of $\Sigma_n$ represents the noise power in each pixel. 
In a typical measurement with e.g. Gaussian error, $\Sigma_n$ is diagonal since errors do not correlate between pixels. 
The Wiener filter, as `(signal) over (noise plus signal)', effectively de-weighs high noise or masked pixels inversely proportional to its signal-to-noise ratio (SNR; zero for masked pixels) and uses the correlation between pixels to `in-paint' the data. 

Ideally, $\Sigma_s$ and $\Sigma_n$ are computed from many repeated measurements. 
In practice, however, we often only have a single measurement of the observed quantities, and have to estimate $\Sigma_s$ and $\Sigma_n$. 
As long as the estimates for $\Sigma_s$ and $\Sigma_n$ are sensible, the resulting $W$ will be a fairly accurate approximation of the true optimal $W$ \citep[Ch. 13.3]{pressNumericalRecipes3rd2007}. 
We present a few typical ways to estimate $\Sigma_s$ and $\Sigma_n$. 

For $\Sigma_n$, if measurement uncertainties are available, those can be taken as a reasonable estimate for the former. 
Alternatively, if the error is known to be white, one can take a Fourier transform of the data vector $d$, and estimate the noise power from averaging high-frequency terms where the signal is far below the noise floor, and noise dominates. 

For $\Sigma_s$, if the process that generate $s$ is known, that can be used to calculate $\Sigma_s$. 
Another way is to simply subtract $\Sigma_n$ from the data covariance $\Sigma_d = \langle d d^\dagger\rangle$ since
\begin{align}
    \langle d d^\dagger\rangle &= \langle (Hs+n) (Hs+n)^\dagger \rangle\nonumber\\
    &= H\langle s s^\dagger \rangle H^\dagger + \langle n n^\dagger \rangle\nonumber\\
    &= H \Sigma_s H^\dagger + \Sigma_n
\end{align}
so that
\begin{align}
    \label{eq:sigman}
    \Sigma_s = H^{-1} (\Sigma_d - \Sigma_n) (H^\dagger)^{-1}.
\end{align}
Substituting equation~(\ref{eq:sigman}) into equation~(\ref{eq:wienerfilter}),
\begin{align}
    W = H^{-1}(\Sigma_d - \Sigma_n)(\Sigma_d)^{-1}.
\end{align}
This can be difficult to computationally evaluate if $H$ is not the identity matrix, or not well defined if $H$ is singular. 
Finally, since the covariance matrix and the power spectrum are related by a Fourier transform, we can compute the power spectrum of $d$ by taking the square modulus of its Fourier transform, model it as a power-law $\propto f^{-\alpha}$, then inverse Fourier transform the fitted power-law curve to get an estimate of the covariance matrix. 

We show our in-painted dynamic spectrum of the first eclipse, with the eclipse masked, in Fig.~\ref{fig:wiener_dynspec}, with $\Sigma_s$ estimated from averaging the power spectra of the full observation according to Welch's method, and inverse Fourier transformed to obtain the covariance matrix, and $\Sigma_n$ taken to be the error on $a(t,\nu)$ from the template matching procedure. 
We note that fitting an anisotropic power-law of the form $f_x^{-\alpha}\cdot f_y^{-\beta}$ to the power spectrum as an estimate for $\Sigma_s$ also works relatively well. Using a power-law not mean that the underlying signal comes from a power-law process, only that a power-law process is a reasonable estimate of the signal power spectrum.
The `raw' dynamic spectrum is shown in the top left-hand panel, with clear sub-band edges, instrumental dropout around time bin $\sim 80$, and various pixels masked from the RFI detection. 
The top right-hand panel shows the Wiener-filtered dynamic spectrum. 
Our boolean mask $H$ is shown in the bottom left-hand panel, with RFI pixels, sub-band edges, bad channels, and bad time bins masked.
The bottom right-hand panel shows the difference between the raw dynamic spectrum and the in-painted dynamic spectrum. 
The masked pixels show the largest differences since the SNR is 0 in those pixels, and the filter takes no information from the pixels themselves. 
Small changes can also be seen in the unmasked region. 

One thing to note is that, naively, the Wiener filter $W$ involves the evaluation of inverses of the covariance matrices which scale as $(\text{number of pixels})^2$, which is computationally inefficient for even moderately sized images. 
However, since the inverse is not explicitly needed, a simple, but substantial speed and memory improvement can be obtained by using iterative conjugate gradient solvers, such as the ones implemented in \texttt{scipy}. 
That is, instead of computing $(H \Sigma_s H^\dagger + \Sigma_n)^{-1}$ explicitly, iteratively solving for $(H \Sigma_s H^\dagger + \Sigma_n)\,x = d$, and then $s_\text{WF} =  \Sigma_s H^{\dagger} x$. 

\end{appendix}

\label{lastpage}

% \end{onecolumn}
\end{document}